\documentclass{article}

\usepackage[english]{babel}

\usepackage[letterpaper,top=2cm,bottom=2cm,left=3cm,right=3cm,marginparwidth=1.75cm]{geometry}

\usepackage{amsmath}
\usepackage{amsfonts}
\usepackage{graphicx}
\usepackage{subcaption}
\usepackage[colorlinks=true, allcolors=blue]{hyperref}
\usepackage[dvipsnames,table]{xcolor}
\usepackage{algorithm}%
\usepackage{algorithmicx}%
\usepackage{algpseudocode}
\usepackage{comment}
\usepackage{multirow}
\usepackage{todonotes}
\usepackage{booktabs}
\usepackage{siunitx}

\usepackage{chemfig}
\setchemfig{atom sep=1.25em}
\definecolor{mygray}{rgb}{0.9, 0.9, 0.9}
\definecolor{myCyan}{HTML}{c4e1e6}
\definecolor{matchyellow}{HTML}{f78c1b}
\definecolor{matchgreen}{HTML}{12e23f}

\title{A Collision-based strategy for Network-free Exploration of Complex Molecular Networks}
\author{Adittya Pal$^2$ \and Rolf Fagerberg$^2$ \and Jakob Lykke Andersen$^2$ \and Peter Dittrich$^1$ \and Daniel Merkle$^{2,3}$}
\date{
    \begin{small}
        $^1$Department of Mathematics and Computer Science, Friedrich Schiller University Jena\\
        $^2$Department of Mathematics and Computer Science, University of Southern Denmark\\
        $^3$Faculty of Technology, Bielefeld University\\[2ex]
    \end{small}
}

\begin{document}
\maketitle

\begin{abstract}

This work presents a stochastic exploration framework for large, implicitly defined chemical reaction spaces that are too large to be generated and stored as explicit molecular networks. 
The exploration strategy mimics stochastic chemical kinetics by combining collision-based pair selection with reaction-template instantiation on demand. 
In each step, the algorithm first samples molecules to collide, then samples a reaction template, and finally samples a concrete reaction instance among the matches of that template. This collision-first factorization avoids exhaustive enumeration of all currently possible reactions and enables exploration of large atomistic reaction spaces under open- or closed-system conditions. We demonstrate the framework on formose chemistry as a case study and analyse both the chemical behaviour reached by the exploration and the computational effects of caching. The implementation 
is intended as a general tool for exploratory analysis of generative reaction systems.

\end{abstract}

\section{Introduction}
Connected systems involving multiple interacting entities are often modelled as networks. An example of such a system is a collection of molecules, which might undergo reactions among themselves. We use the term \emph{molecular network} (or \emph{chemical reaction network}) when modelling a set of molecules along with the possible reactions between them.
In such representations of the system as a network, molecules constitute the nodes and potential reactions define the edges~\cite{abstract_graph_transform}.
Molecular networks provide a convenient abstraction for querying the system how particular target molecules may arise from given source molecules, i.e., for finding pathways in the network.

An often-used approach to solve such queries is to consider them as a problem of transport through that network and then use either path-finding algorithms~\cite{bfs_crn,bfs_crn2,dfs_crn,dfs_crn2} or methods based on network flow modelling~\cite{jakob_integer_hyperflows,thermoflow,thermoFlow2}. These methods assume that the full molecular network is either available at the beginning or is created during the query process.
However, for chemically rich systems, such as the formose reaction~\cite{complex_formose,complex_formose2}, constructing, storing, and querying the full molecular network often become increasingly difficult as its size and density increases due to combinatorial growth in the number of molecules and reactions~\cite{large_crn,large_crn2}.
Also, large portions of the full network may be irrelevant, e.g.\ when specific starting conditions for the system lead it to populate only subsets of the potential chemical space.

Ab initio methods~\cite{abinitio1,abinitio2,abinitio3} and molecular dynamics~\cite{md1,md2,md3} offer alternative approaches to pathway discovery that avoids explicit network construction. By integrating atomic trajectories directly, it is possible to record the sequence of molecular transformations as they occur, constructing the pathway according to the equations governing the system.
Yet these simulations remain restricted to relatively small systems and short timescales due to their computation-intensive nature, and therefore cannot capture the expansive chemical spaces or extended reaction cascades characteristic of e.g.\ prebiotic or polymerization chemistry.~\cite{highSimCost1,highSimCost2}

Stochastic simulation methods provide a middle ground between these two extremes. 
By sampling reaction events probabilistically, they enable the exploration of chemical state spaces that are orders of magnitude larger than those accessible to deterministic or fully atomistic approaches for extended timescales. 
Such methods also reflect the inherently discrete and stochastic nature of chemical reactions, particularly in low-concentration or prebiotic contexts where fluctuations play an essential role. 
Classical stochastic methods, including kinetic Monte Carlo~\cite{kmc} and the Gillespie algorithm~\cite{gillespie,gillespie2,gillespie3}, used in tools like uch as NFsim~\cite{nfsim}, Kappa~\cite{kappa}, and BioNetGen~\cite{bngl,bngl2},  
have been successfully applied in systems biology, catalysis, and polymerization to explore molecular networks, trace pathways and trace the evolution of the system over time.
Nonetheless, traditional variants typically require explicit enumeration of all possible reactions at each simulation step—a requirement that may become limiting when the reaction space is combinatorially large.

One key aspect of the methodology presented in this paper is to avoid such enumeration in each exploration step of all currently possible reactions
in order to sample from those. 
Our approach is instead \emph{first} sample molecules to collide, and \emph{then} consider which reactions they can take part in.

Therefore, in this work, we introduce a stochastic exploration algorithm for large, implicit molecular networks generated by repeated application of reaction templates. The goal is to sample dynamically reachable regions of chemical space without prior network enumeration. To that end, we use a discrete-time Markov process whose transition policy is intentionally factorized into three stages: sampling a colliding molecular pair, sampling a reaction template, and sampling a reaction instance induced by that template on the chosen pair. This induces a specific bias over reachable reactions and states: transitions are favoured by molecular abundance through collision sampling, while templates and instances are sampled uniformly unless the user provides alternative weights. The resulting process should therefore be understood as a principled exploration policy rather than a kinetically faithful stochastic simulator. Its utility lies in reachability, discoverability, and controllable traversal of large reaction spaces where exhaustive network construction or reliable kinetic parametrization is infeasible.

\section{Methods}

In this section, we outline the stochastic process employed in the exploration of chemical systems in this work.

\subsection{Collisions as a Basis for Reactions}
\label{collision_theory}

At the molecular scale, chemical reactions are probabilistic events that arise from collisions between molecules. Hence, dynamics of chemical systems are inherently stochastic at the molecular level. Unlike deterministic macroscopic models, such as systems of differential equations, stochastic simulation directly model the discrete and random nature of chemical reactions. We here briefly recap the collision theory of chemical kinetics, as it provides a physics-based framework for understanding how molecules collide, interact, and react into products.

According to the collision theory~\cite{collisionTheory}, a reaction occurs when molecules collide with sufficient kinetic energy to overcome an activation energy barrier and with a relative orientation that allows bond rearrangement. The rate of a reaction~$k_{12}$ between two molecules $m_1$ and $m_2$ is given by: 

\begin{equation*}
    k_{12} = z_{12}\cdot p_{12}\, e^{\frac{-E_a}{RT}}
    \label{rate}
\end{equation*}
where
\begin{center}
\begin{tabular}{rl}
      $z_{12}$ & is the collision frequency between the molecules $m_1$ and $m_2$, \\
     $p_{12}$ & is the probability that a collision between the molecules $m_1$ and $m_2$ \\
     & leads to a reaction, and\\
     $e^{\frac{-E_a}{RT}}$ & is the scaling of the rate of the reaction with the temperature of the system, \\
     & which is a measure of the average kinetic energy of the molecules in the system.
\end{tabular}
\end{center}
The collision frequency $z_{12}$ is derived from the kinetic theory of gases and depends on the physical properties of the molecules and the system. For an ideal gas, it may be approximated as:

\begin{equation*}
    z_{12} = \sigma_{12} \sqrt{\frac{8k_BT}{\pi\mu_{12}}} [m_1][m_2]
\end{equation*}
where
\begin{center}
\begin{tabular}{rl}
     $\sigma_{12}$ & is the effective cross-section for collision of the molecules $m_1$ and $m_2$, \\
     $k_B$ & is the Boltzmann constant, \\
     $T$ & is the temperature of the system in Kelvin, \\
     $\mu_{12}$ & is the reduced mass of the molecular pair $(m_1, m_2)$, and\\
     $[m_1]$, $[m_2]$ & are the concentrations of the molecules $m_1$ and $m_2$, respectively.  
\end{tabular}
\end{center}
We note that assuming a fixed volume for the system, the concentration $[m_i]$ is proportional to the number $N(m_i)$ of molecules~$m_i$ in the system.
We will formulate our methodology using the latter.
This also better reflects the quantization effect on the concentration values possible which low-concentration settings have.

While unimolecular reactions might appear to not fit into a scheme with a pair of colliding molecules, theories for unimolecular reactions~\cite{unimolecularReaction} consider the first step of such a reaction to be the energization of the reacting molecule by a collision with another molecule. Therefore, collisions can still be considered to act as a trigger for unimolecular reactions.

\subsection{Effecting a Reaction}
\label{reaction_enumerate}

As discussed in the previous section, collision theory considers that for a reaction to occur, the molecules involved must first collide with a sufficient energy and an appropriate orientation. Once such a collision is registered, the next step is to determine the possible reactions that might occur.

\begin{figure}
    \centering
    \begin{subfigure}{\textwidth}
        \centering
        \begin{tikzpicture}[scale=1.0]    
        \tikzset{
            arrow/.style = {->, -{Stealth[length = 5pt]}},
            bond/.style={line width=0.6pt},
            dbond/.style={double, double distance=1pt}
        }
        \draw[fill=lime,fill opacity=0.1,draw=none] (0,-0.25) rectangle (1,1.25);
        \node at (-0.2, 1) {\scalebox{0.8}{$L$}};
        \node (c0) at (0.75,0.5) {\tiny{C}};
        \node (o1) at (0.75,1) {\tiny{O}};
        \node (c2) at (0.25,0.3) {\tiny{C}};
        \node (h3) at (0.25,0) {\tiny{H}};
        \draw[bond] (c2) to (h3);
        \draw[bond] (c0) to (c2);
        \draw[dbond] (c0) to (o1);
        
        \draw[arrow] (1,0.5) to (4,0.5);
        \node at (2.5,0.7) {\tiny{replace by}};

        \draw[fill=lime,fill opacity=0.1,draw=none] (4,-0.25) rectangle (5,1.25);
        \node at (5.2, 1) {\scalebox{0.8}{$R$}};
        \node (c0) at (4.75,0.3) {\tiny{C}};
        \node (o1) at (4.75,0.8) {\tiny{O}};
        \node (c2) at (4.25,0.1) {\tiny{C}};
        \node (h3) at (4.75,1) {\tiny{H}};
        \draw[dbond] (c0) to (c2);
        \draw[bond] (c0) to (o1);

        \draw[fill=cyan,fill opacity=0.1,draw=none] (0,-2) rectangle (1.7,-1);
        \begin{scope}[rotate around={30:(0,-2)}]
            \draw[rounded corners,fill=myCyan,draw=none] (0.6, -2.3) rectangle (1.2, -2.05);
        \end{scope}
        \draw[rounded corners,fill=myCyan,draw=none] (0.86, -1.7) rectangle (1.14, -1.1);
        \node (lhs) at (0.6,-1.5) {\tiny{\chemfig{HO-[:-30]-[:30]=[:90]O}}};
        \node (lhs2) at (1.5,-1.5) {\tiny{\chemfig{=[:90]O}}};
        \node at (-0.2,-1.8) {\scalebox{0.8}{$G$}};
        \draw[fill=cyan,fill opacity=0.1,draw=none] (3.3,-2) rectangle (5,-1);
        \begin{scope}[rotate around={30:(3.3,-2)}]
            \draw[rounded corners,fill=myCyan,draw=none] (3.9, -2.3) rectangle (4.5, -2.05);
        \end{scope}
        \draw[rounded corners,fill=myCyan,draw=none] (4.16, -1.7) rectangle (4.44, -1.12);
        \draw[rounded corners,fill=myCyan,draw=none] (4.2, -1.12) rectangle (4.65, -1.37);
        \node (rhs) at (4,-1.5) {\tiny{\chemfig{HO-[:-30]=[:30]-[:90]OH}}};        
        \node (rhs2) at (4.8,-1.5) {\tiny{\chemfig{=[:90]O}}};
        \node at (5.2,-1.8) {\scalebox{0.8}{$H$}};
        \draw[arrow] (1.7,-1.5) to (3.3,-1.5);
        \node at (2.5,-1.3) {\tiny{transform to}};
        
        \draw[arrow] (0.5,-0.25) to (0.5,-1);
        \node at (1,-0.65) {\tiny{matches}};
        \draw[arrow] (4.5,-0.25) to (4.5,-1);
        \node at (4,-0.65) {\tiny{matches}};        
\end{tikzpicture}
        \caption{Application of a template to a single graph in the pair leading to a keto-enol tautomerism isomerisation.}
        \label{fig:rule1}
    \end{subfigure}
    \\
    \vspace{0.1cm}
    \begin{subfigure}{\textwidth}
        \centering
        \begin{tikzpicture}[scale=1.0]    
        \tikzset{
            arrow/.style = {->, -{Stealth[length = 5pt]}},
            bond/.style={line width=0.6pt},
            dbond/.style={double, double distance=1pt}
        }
        \draw[fill=lime,fill opacity=0.1,draw=none] (0,-0.25) rectangle (1.5,1.25);
        \node at (-0.2, 1) {\scalebox{0.8}{$L$}};
        \node (c0) at (1.25,0.3) {\tiny{C}};
        \node (o1) at (1.25,0.8) {\tiny{O}};
        \node (c2) at (0.75,0.1) {\tiny{C}};
        \node (c4) at (0.25,0.3) {\tiny{C}};
        \node (o5) at (0.25,0.8) {\tiny{O}};
        \node (h3) at (0.25,1) {\tiny{H}};
        \draw[bond] (c2) to (c4);
        \draw[bond] (c0) to (c2);
        \draw[dbond] (c0) to (o1);
        \draw[bond] (c4) to (o5);
        
        \draw[arrow] (1.5,0.5) to (3.5,0.5);
        \node at (2.5,0.7) {\tiny{replace by}};

        \draw[fill=lime,fill opacity=0.1,draw=none] (3.5,-0.25) rectangle (5,1.25);
        \node at (5.2, 1) {\scalebox{0.8}{$R$}};
        \node (c0) at (4.75,0.3) {\tiny{C}};
        \node (o1) at (4.75,0.8) {\tiny{O}};
        \node (c2) at (4.25,0.1) {\tiny{C}};
        \node (h3) at (4.75,1) {\tiny{H}};
        \node (c4) at (3.75,0.3) {\tiny{C}};
        \node (o5) at (3.75,0.8) {\tiny{O}};
        \draw[dbond] (c0) to (c2);
        \draw[bond] (c0) to (o1);
        \draw[dbond] (c4) to (o5);

        \draw[fill=cyan,fill opacity=0.1,draw=none] (0,-2.3) rectangle (1.5,-1);
        \begin{scope}[rotate around={30:(0,-2)}]
            \draw[rounded corners,fill=myCyan,draw=none] (0.65, -2.3) rectangle (1.2, -2.05);
        \end{scope}
        \begin{scope}[rotate around={-30:(0,-2)}]
            \draw[rounded corners,fill=myCyan,draw=none] (0.1, -1.3) rectangle (0.6, -1.55);
        \end{scope}
        \draw[rounded corners,fill=myCyan,draw=none] (0.91, -1.7) rectangle (1.19, -1.1);
        \draw[rounded corners,fill=myCyan,draw=none] (0.31, -1.7) rectangle (0.59, -1.1);
        \draw[rounded corners,fill=myCyan,draw=none] (0.35, -1.1) rectangle (0.75, -1.4);
        \node (rhs) at (0.75,-1.7) {\tiny{\chemfig{HO-[:-90]-[:-30](-[:-90]OH)-[:30]=[:90]O}}}; 
        \node at (-0.2,-2.1) {\scalebox{0.8}{$G$}};
        \draw[fill=cyan,fill opacity=0.1,draw=none] (3.5,-2.3) rectangle (5,-1);
        \begin{scope}[rotate around={30:(3.5,-2)}]
            \draw[rounded corners,fill=myCyan,draw=none] (4.12, -2.35) rectangle (4.72, -2.05);
        \end{scope}
        \draw[rounded corners,fill=myCyan,draw=none] (4.42, -1.7) rectangle (4.7, -1.1);
        \draw[rounded corners,fill=myCyan,draw=none] (4.5, -1.1) rectangle (4.88, -1.4);
        \draw[rounded corners,fill=myCyan,draw=none] (3.65, -1.7) rectangle (3.95, -1.1);
        \node (lhs) at (4.5,-1.7) {\tiny{\chemfig{HO-[:90]=[:30]-[:90]OH}}};
        \node (lhs2) at (3.8,-1.4) {\tiny{\chemfig{=[:90]O}}};
        \node at (5.2,-2.1) {\scalebox{0.8}{$H$}};
        \draw[arrow] (1.5,-1.7) to (3.5,-1.7);
        \node at (2.5,-1.5) {\tiny{transform to}};
        
        \draw[arrow] (0.75,-0.25) to (0.75,-1);
        \node at (1.25,-0.65) {\tiny{matches}};
        \draw[arrow] (4.25,-0.25) to (4.25,-1);
        \node at (3.75,-0.65) {\tiny{matches}};        
\end{tikzpicture}
        \caption{Application of a template to a single graph from the collision pair leading to its dissociation.}
        \label{fig:rule2}
    \end{subfigure}
    \\
    \vspace{0.1cm}
    \begin{subfigure}{\textwidth}
        \centering
        \begin{tikzpicture}[scale=1.0]    
        \tikzset{
            arrow/.style = {->, -{Stealth[length = 5pt]}},
            bond/.style={line width=0.6pt},
            dbond/.style={double, double distance=1pt}
        }
        \draw[fill=lime,fill opacity=0.1,draw=none] (0,-0.25) rectangle (1.5,1.25);
        \node at (-0.2, 0.5) {\scalebox{0.8}{$L$}};
        \node (c0) at (1.25,0.3) {\tiny{C}};
        \node (o1) at (1.25,0.8) {\tiny{O}};
        \node (c2) at (0.75,0.1) {\tiny{C}};
        \node (h3) at (1.25,1) {\tiny{H}};
        \node (c4) at (0.25,0.3) {\tiny{C}};
        \node (o5) at (0.25,0.8) {\tiny{O}};
        \draw[dbond] (c0) to (c2);
        \draw[bond] (c0) to (o1);
        \draw[dbond] (c4) to (o5);
        
        \draw[arrow] (1.5,0.5) to (3.5,0.5);
        \node at (2.5,0.7) {\tiny{replace by}};

        \draw[fill=lime,fill opacity=0.1,draw=none] (3.5,-0.25) rectangle (5,1.25);
        \node at (5.2, 0.5) {\scalebox{0.8}{$R$}};
        \node (c0) at (4.75,0.3) {\tiny{C}};
        \node (o1) at (4.75,0.8) {\tiny{O}};
        \node (c2) at (4.25,0.1) {\tiny{C}};
        \node (c4) at (3.75,0.3) {\tiny{C}};
        \node (o5) at (3.75,0.8) {\tiny{O}};
        \node (h3) at (3.75,1) {\tiny{H}};
        \draw[bond] (c2) to (c4);
        \draw[bond] (c0) to (c2);
        \draw[dbond] (c0) to (o1);
        \draw[bond] (c4) to (o5);

        \draw[fill=cyan,fill opacity=0.1,draw=none] (0,-2) rectangle (1.5,-0.7);
        \begin{scope}[rotate around={30:(0,-2)}]
            \draw[rounded corners,fill=myCyan,draw=none] (0.6, -1.97) rectangle (1.15, -1.67);
        \end{scope}
        \draw[rounded corners,fill=myCyan,draw=none] (0.7, -1.4) rectangle (0.95, -0.82);
        \draw[rounded corners,fill=myCyan,draw=none] (0.75, -0.81) rectangle (1.15, -1.11);
        \draw[rounded corners,fill=myCyan,draw=none] (1.1, -1.8) rectangle (1.4, -1.2);
        \node (rhs) at (0.6,-1.4) {\tiny{\chemfig{HO-[:-90]-[:-30](-[:-90]OH)=[:30]-[:90]OH}}}; 
        \node (lhs2) at (1.25,-1.5) {\tiny{\chemfig{=[:90]O}}};
        \node at (-0.2,-1.8) {\scalebox{0.8}{$G^\prime$}};
        \draw[fill=cyan,fill opacity=0.1,draw=none] (3.5,-2) rectangle (5,-0.7);
        \begin{scope}[rotate around={30:(3.5,-2)}]
            \draw[rounded corners,fill=myCyan,draw=none] (4.17, -1.98) rectangle (4.72, -1.72);
            \draw[rounded corners,fill=myCyan,draw=none] (4.17, -2.3) rectangle (4.77, -2.05);
            \draw[rounded corners,fill=myCyan,draw=none] (4.17, -2.3) rectangle (4.37, -1.72);
        \end{scope}
        \draw[rounded corners,fill=myCyan,draw=none] (4.26, -1.4) rectangle (4.54, -0.8);
        \draw[rounded corners,fill=myCyan,draw=none] (4.5, -1.47) rectangle (4.9, -1.73);        
        \node (lhs) at (4.25,-1.35) {\tiny{\chemfig{HO-[:-90]-[:-30](-[:-120]HO)(-[:-60]-[:30]OH)-[:30]=[:90]O}}};    
        \node at (5.2,-1.8) {\scalebox{0.8}{$H^\prime$}};
        \draw[arrow] (1.5,-1.35) to (3.5,-1.35);
        \node at (2.5,-1.5) {\tiny{transform to}};

        \draw[fill=cyan,fill opacity=0.1,draw=none] (0,1.8) rectangle (1.5,3.1);
        \begin{scope}[rotate around={30:(0,1.8)}]
            \draw[rounded corners,fill=myCyan,draw=none] (0.6, 1.87) rectangle (1.15, 2.17);
        \end{scope}
        \draw[rounded corners,fill=myCyan,draw=none] (0.38, 1.9) rectangle (0.65, 2.4);
        \draw[rounded corners,fill=myCyan,draw=none] (0.4, 1.88) rectangle (0.85, 2.13);
        \draw[rounded corners,fill=myCyan,draw=none] (1.1, 2) rectangle (1.4, 2.6);
        \node (rhs) at (0.6,2.45) {\tiny{\chemfig{HO-[:-90]-[:-30](-[:-90]OH)=[:30]-[:90]OH}}}; 
        \node (lhs2) at (1.25,2.3) {\tiny{\chemfig{=[:90]O}}};
        \node at (-0.2,2.9) {\scalebox{0.8}{$G$}};
        \draw[fill=cyan,fill opacity=0.1,draw=none] (3.5,1.8) rectangle (5,3.1);
        \begin{scope}[rotate around={30:(3.5,1.8)}]
            \draw[rounded corners,fill=myCyan,draw=none] (4.1, 1.9) rectangle (4.65, 2.15);
        \end{scope}
        \begin{scope}[rotate around={-30:(3.5,1.8)}]
            \draw[rounded corners,fill=myCyan,draw=none] (3.75, 2.75) rectangle (4.3, 2.95);
        \end{scope}
        \draw[rounded corners,fill=myCyan,draw=none] (3.87, 1.87) rectangle (4.15, 2.45);
        \draw[rounded corners,fill=myCyan,draw=none] (4.5, 1.88) rectangle (4.95, 2.13);
        \draw[rounded corners,fill=myCyan,draw=none] (4.5, 1.87) rectangle (4.75, 2.45);
        \node (lhs) at (4.25,2.45) {\tiny{\chemfig{HO-[:-90]-[:-30](=[:-90]O)-[:30](-[:90]OH)-[:-30]-[:-90]OH}}};
        
        \node at (5.2,2.9) {\scalebox{0.8}{$H$}};
        \draw[arrow] (1.5,2.45) to (3.5,2.45);
        \node at (2.5,2.6) {\tiny{transform to}};
        
        \draw[arrow] (0.75,-0.25) to (0.75,-0.7);
        \node at (1.25,-0.43) {\tiny{matches}};
        \draw[arrow] (4.25,-0.25) to (4.25,-0.7);
        \node at (3.75,-0.43) {\tiny{matches}};   
        \draw[arrow] (0.75,1.25) to (0.75,1.8);
        \node at (1.25,1.53) {\tiny{matches}};
        \draw[arrow] (4.25,1.25) to (4.25,1.8);
        \node at (3.75,1.53) {\tiny{matches}}; 
\end{tikzpicture}
        \caption{Two possible matches of the template with the union graph of the pair of molecules leading to two possible aldol addition reactions.}
        \label{fig:rule3}
    \end{subfigure}
    \\
    \vspace{0.1cm}
    \begin{subfigure}{\textwidth}
        \centering
        \begin{tikzpicture}[scale=1.2]    
            \tikzset{
                arrow/.style = {->, -{Stealth[length = 5pt]}},
                bond/.style={line width=0.6pt},
                dbond/.style={double, double distance=1pt},
                atomNode/.style={}
            }
            
            \draw[fill=lime,fill opacity=0.1,draw=none] (0,-0.2) rectangle (1.15,0.9);
            \node at (-0.2, 0.7) {\scalebox{0.8}{$L$}};
            \node (c0) at (0.75,0.2) {\scalebox{0.5}{C}};
            \node (o1) at (0.75,0.7) {\scalebox{0.5}{O}};
            \node (c2) at (0.25,0) {\scalebox{0.5}{C}};
            \node (h3) at (0.9,0.7) {\scalebox{0.5}{H}};
            \draw[dbond] (c0) to (c2);
            \draw[bond] (c0) to (o1);        
                
            \draw[arrow] (1.15,0.35) to (3.85,0.35);
            \node at (2.5,0.5) {\tiny{replace by}};
        
            \draw[fill=lime,fill opacity=0.1,draw=none] (3.85,-0.2) rectangle (5,0.9);
            \node at (5.2, 0.7) {\scalebox{0.8}{$R$}};
            \node (c0) at (4.75,0.2) {\scalebox{0.5}{C}};
            \node (o1) at (4.75,0.7) {\scalebox{0.5}{O}};
            \node (c2) at (4.25,0) {\scalebox{0.5}{C}};
            \node (h3) at (4.1,0) {\scalebox{0.5}{H}};
            \draw[bond] (c0) to (c2);
            \draw[dbond] (c0) to (o1);
        
            \draw[fill=cyan,fill opacity=0.1,draw=none] (0,-1.2) rectangle (1.7,-0.4);
            \node (lhs) at (0.6,-0.8) {\tiny{\chemfig{HO-[:-30]-[:30]=[:90]O}}};
            \node (lhs2) at (1.5,-0.8) {\tiny{\chemfig{=[:90]O}}};
            \node at (-0.2,-1) {\scalebox{0.8}{$G$}};
            \node at (2.8,-0.95) {\textcolor{red}{\tiny{no reaction possible}}};
                
            \node at (1.1,-0.25) {\textcolor{red}{\tiny{no match}}};
        \end{tikzpicture}
        \caption{No isomerization reaction is possible if the left side of the template is not a subgraph of the molecules.}
        \label{fig:rule4}
    \end{subfigure}
    \caption{Instantiation of reactions using the reaction template and the chosen pair of molecules.}
    \label{fig:reactionTemplate}
\end{figure}
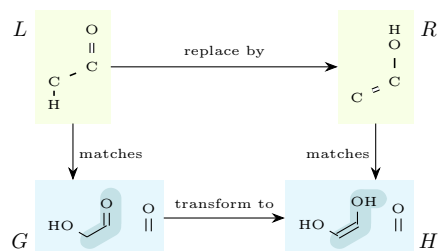
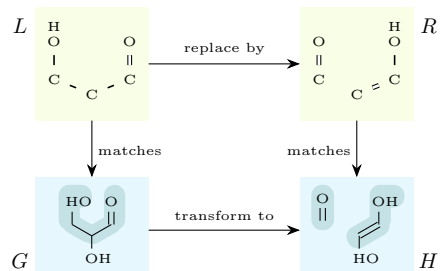
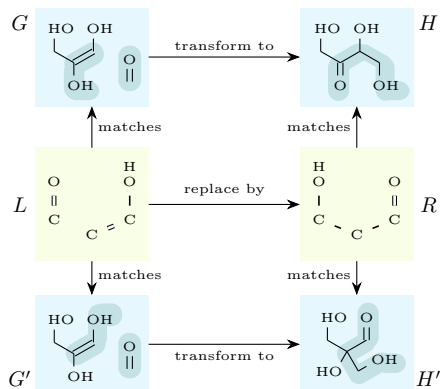
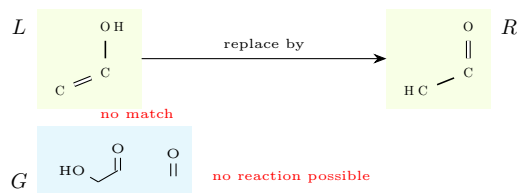

In complex chemical systems, direct enumeration of all possible reactions between a colliding pair of molecules might be computationally expensive, and can lead to an intractable growth of the molecular space.
Instead, we take advantage of the observation that often an entire class of reactions have identical change in the bond set, and employ a two-step reaction selection procedure based on \emph{reaction templates}.
A reaction template is a specification of such a common change in bond set and of the atoms involved in it. This can be seen as modelling a \emph{class} of reactions, namely those reactions whose actions agree with the specification.
As an example, in the formose system~\cite{formose,formose2}, the relevant reaction classes are keto-enol tautomerism, aldol condensation, and their reverse directions---thus, this system is well described by only two reaction templates and their inverses.
If molecules are represented as molecular graphs, then reaction templates are naturally represented as graph transformation rules~\cite{jakob_graph_grammar}. A graph transformation rule may match a given molecular graph, in which case the specified transformation implies a change into another molecular graph---in other words, the rule has been instantiated to a specific reaction. See Figure~\ref{fig:reactionTemplate} for concrete examples. Computational frameworks such as MØD~\cite{mod_graphTransformation} exist which make this notion executable.
Thus, an entire chemical network can be defined and generated in silico by supplying a set of reaction templates and a set of initial molecular graphs\footnote{Such two sets are sometimes denoted a \emph{graph grammar}.} and then repeatedly applying the reaction templates to the currently available molecular graphs.

We now describe how reaction templates are employed in our methodology. As a first step, a reaction template $r$ is chosen randomly from the given library $R$ of reaction templates relevant to the system.
This can be considered as selecting the class of reactions that might occur to the colliding pair of molecules.

In the second step, the selected template $r\in R$ is applied to the colliding molecules $(m_1, m_2)$ as shown in Figure~\ref{fig:reactionTemplate}.
The application of the reaction template $r$ to the chosen pair $(m_1, m_2)$ requires a match (formally, a graph monomorphism) to exist
with the left side of the template. If a match is found between the fragment on the left side of $r$ and $(m_1, m_2)$, i.e., the graph on the left side of the template is a subgraph of the union $m_1 \cup m_2$ of the two graphs $m_1$ and $m_2$, then the type of reaction represented by $r$ can occur with $(m_1, m_2)$. 
A reaction is then instantiated by applying the chosen template $r$ to the molecular pair $(m_1, m_2)$ to generate the product molecules.

It might be possible that there are multiple matches of the reaction template $r$ in the colliding pair $(m_1, m_2)$. In that case, each instantiation corresponds to a different reaction. A single reaction can be selected from all reactions effected by the template on the pair of molecules uniformly at random or weighted by the relative probabilities of each reaction.
The relative probability of a reaction with rate constant $k$ can be modelled using either the thermodynamic free energies or the kinetics (the energy barriers) of the reactions. For a time interval $\Delta t$, the probability of reaction is roughly $k\times \Delta t$. The calculation of rate constants $k$ for each reaction is computationally expensive and often not practically feasible. Therefore, instead of inaccurate estimates, we choose to use a uniform probability distribution both while selecting a reaction template $r$ (from the library of templates $R$) and while selecting the particular reaction instance from the reactions possible on applying template $r$ to the chosen colliding molecular pair $(m_1, m_2)$. 
This is a deliberate design choice: in chemically rich systems such as prebiotic networks, reliable kinetic parameters are often unknown or highly uncertain,
and introducing guessed rate constants risks biasing the exploration.
Uniform sampling avoids any particular bias and prioritizes discoverability of structurally reachable regions of the implicit reaction network defined by the reaction templates. The resulting stochastic process is intended as a generic exploratory walk through chemical space rather than an approximation of physical reaction kinetics. However, the framework allows the user to make alternative custom sampling choices, either based on actual chemical knowledge, or, in the other direction, use artificial values to direct the exploration of the chemical space in specific directions.

If the graph on the left side of the reaction template is not a subgraph of the union graph $m_1 \cup m_2$, there is no match and the reaction template cannot be applied to the molecular pair $(m_1, m_2)$. No reaction occurs in this case.

This two-step approach offers two principal advantages.
First, it reduces computational cost by deferring structural enumeration until a reaction template is chosen,
thereby avoiding exhaustive generation of all possible reactions.
Second, it ensures mechanistic diversity: templates corresponding to less frequent but chemically important reaction mechanisms can,
if desired, remain accessible in the simulation by an appropriate choice of parameters, rather than being overshadowed by mechanisms with numerous instantiations.

\subsection{Towards a Stochastic Process}
\label{stocProcess}

We choose to represent the state~$s[i]$ of the chemical system at the simulation step $i$ by a vector of molecule counts. Since we do not want to enumerate all molecules beforehand, we store the state as a mapping (in practice, a hash map) of the non-zero counts, with molecules as keys and counts as values. If the state at step $i$ has $N$ molecules with non-zero counts, then the state can be visualized as
\begin{equation*}
    s[i] = \begin{bmatrix} m_1\colon N(m_1) \\ m_2\colon N(m_2) \\ \vdots \\ m_N\colon N(m_N) \end{bmatrix}
\end{equation*}
where $N(m_i)$ is the count of the molecule $m_i$. Below, we specify changes to this state using vector notation, with the natural meaning. The likelihood of choosing a given pair of molecules for collision depends on the number of pairs of that type. For example, the probability of a pair of molecules of different classes $m_1$ and $m_2$ colliding would be
\begin{equation}
    \mathcal{P}\left((m_1, m_2)\mid s[i]\right) \propto \frac{N(m_1)\cdot N(m_2)}{N (N-1)/2}
    \label{react_prob_different_molecules}
\end{equation}
and the probability of two molecules of class $m_1$ colliding would be
\begin{equation}
  \mathcal{P}\left((m_1, m_1)\mid s[i]\right) \propto \frac{N(m_1)\cdot(N(m_1)-1)/2}{N (N-1)/2}
      \label{react_prob_same_molecules}
\end{equation}
The choice of molecules is made without replacement because a molecule cannot collide with itself. As described in Section~\ref{collision_theory}, other factors influencing these probabilities would be the reduced mass of the molecules $\mu_{12}$ and the effective cross-section of the molecules $\sigma_{12}$ (while the temperature $T$ can be assumed to be the same for all molecules in a well stirred system). Again, estimation of these values may be hard to get by, and since the goal is flexible exploration of chemical spaces, we choose in this work to set them to unity for all reactions, while a user in other settings is free to supply such values to the framework.

Reactions can occur as a result of collisions. Each reaction $j$ effected as explained in Section~\ref{reaction_enumerate} can be described by:
\begin{itemize}
    \item An input stoichiometric vector $\nu_j^-$, which indicates how many molecules each of the $N$ classes are used when the reaction $j$ occurs.
    \item An output stoichiometric vector $\nu_j^+$, which indicates how many molecules of each class is produced when the reaction $j$ occurs. A reaction often produces molecules from a new class.
    \item Any additional information available from thermodynamics or kinetics on how likely reaction $j$ is under the given physical conditions, which can be used to decide if $j$ eventually occurs.
\end{itemize}
An often used term is the stoichiometric change vector~$\nu_j$ which is related to the above as $\nu_j = \nu_j^+ - \nu_j^-$. However, we choose to retain the decomposition into input and output vectors because it allows explicit representation of a molecule as a catalyst if it is present both in the input vector $\nu^-$ and the output vector $\nu^+$.

When a reaction $j$ occurs, the state is updated as:
\begin{equation*}
    s[i+1] \leftarrow s[i] - \nu_j^- + \nu_j^+ 
\end{equation*}
This state update rule is governed by the collision frequency and the stoichiometric vectors of the chosen reaction. Therefore the evolution of the system can be described as a stochastic process. The reader might note a contrast to the Gillespie stochastic algorithm where the time advanced is dependent on the `total reactivity' of the system. Here, it is always advanced by a discrete step. This measure would be better called the simulation time, measured in discrete event steps. It should be interpreted as an ordering parameter rather than physical time. As we describe in Section~\ref{sec:algorithm}, it may be scaled by the number of molecular pairs in the system which affects the collision rate to give a measure of the physical time elapsed since the start of the simulation.

The update step depends on the current state $s[i]$ only and no information from the sequence of past states $s[i-1]$, $s[i-2]$, \ldots influences it. Moreover, the reaction templates and the collision pair selection procedure does not change over time in this framework.

\subsubsection{Adding inflows of source molecules and outflows}
\label{sec:inflowOutflow}
While inflow–outflow dynamics are well established in biological simulators, these frameworks operate at levels of abstraction of sites that preclude explicit atomic bond rearrangements. Conversely, atomistic molecular network exploration methods rarely incorporate stochastic open-system dynamics during simulation.
Therefore, we add two mechanisms to make the exploration closer to that in an open system exchanging mass with the surroundings as in terrestrial systems and in flow reactors:

\begin{enumerate}
    \item \emph{Fixed inflow}: A fixed non-negative vector $\delta$ is added to the current state of the system, if an inflow step is chosen to be effected as the next step in the simulation. Thus, the inflow is modelled as a zero-order reaction independent of the current state $s[i]$, which captures settings where molecules abundant in the surroundings diffuse into the system, as well as flow reactor settings.
    
    \item \emph{Concentration-dependent outflow}: A vector $\epsilon$, assembled proportionally to the counts of individual molecular classes, is subtracted from the state vector, if an outflow step is chosen to be effected as the next step in the simulation. The counts of the molecules of each class in the outflow depend on the composition of the system: in a well-stirred continuous flow reactor or homogenous phase reaction system, molecular classes with higher counts are expected to be present in larger quantities in the outflow too. Due to this dependence of the composition of the outflow vector $\epsilon$ on the current concentration of molecular classes in the system, the outflow is a first order process.   
\end{enumerate}

A convenient choice to decide when to effect an inflow, an outflow, or a reaction to the system is to calculate the relative probabilities of a zero, first and second order reaction. The order of the reaction to be performed is chosen based on these calculated weights. If a zero order reaction is decided to be executed, an inflow step is executed, else if a first order reaction is decided on, an outflow step is executed. When a second order reaction is chosen, a collision between two molecules is simulated by selecting a pair of molecules based on Equation~\eqref{react_prob_different_molecules} and ~\eqref{react_prob_same_molecules}. This might lead to a reaction. Recall from Section~\ref{collision_theory} that we choose to treat unimolecular reactions (like dissociation or isomerisation of molecules) as second order reactions too, as they are postulated to be the result of a collision between the reacting molecule with another molecule.

\begin{figure}
    \centering
    \includegraphics[width=0.5\linewidth]{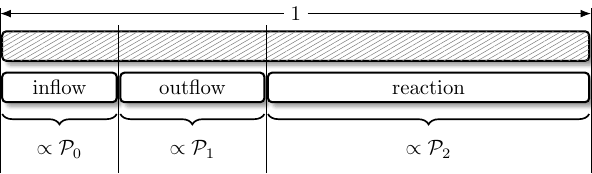}
    \caption{The calculation of the weights for choosing whether to effect an inflow, outflow or a reaction at a particular simulation step.}
    \label{fig:rel_prob}
\end{figure}

Hence, in each simulation step, either a zero order reaction, a first order reaction, or a second order reaction occurs. The relative weights of the different order reactions at a particular simulation step $t$ are calculated as:
\begin{align}
    \mathcal{P}_0 &= k_0 \label{zero_order}\\
    \mathcal{P}_1 &= k_1 \cdot \left(\sum_{i=0}^N n(m_i)\right) \label{first_order}\\
    \mathcal{P}_2 &= k_2 \cdot \frac{\left(\sum_{i=0}^N n(m_i)\right)\left(\left(\sum_{i=0}^N n(m_i)\right) - 1 \right)}{2} \label{second_order}
\end{align}
where $k_0$, $k_1$ and $k_2$ are the stochastic reaction constants for the different order reactions which are user defined parameters of the simulation. They adjust the frequencies of the inflow and outflow steps relative to the rate of collisions.
Equations~\eqref{zero_order}, \eqref{first_order} and \eqref{second_order} stipulate the total number of possible reactions of each order in the system at the simulation step $i$. The weights according to which the order of the reaction is chosen are then calculated by normalizing the results of Equations~\eqref{zero_order}, \eqref{first_order} and \eqref{second_order}. The order of the reaction to be performed is then decided by drawing a random number and seeing what range it falls into, as shown in Figure~\ref{fig:rel_prob}.

The possibility of mass flow in and out of the system not only allows for simulation of open systems, but also permits a given system to be studied under two possible conditions. Without mass flows, the system will converge towards a steady-state, in which the counts of the molecules of each class (which in turn determine the reactions occurring in the system, as discussed in Section~\ref{stocProcess}) make the number of molecules of each class produced by reactions balance the number consumed by reactions. This is analogous to the notion of equilibrium control~\cite{thermodynamic_control}. On the other hand, with mass flows the number of reactions required to form a molecule of given class becomes a major influence on its count in the system, since a longer pathway allows for more outflow of mass on which the molecule class depends. This is analogous to kinetic control~\cite{kinetic_control} in chemical systems.

\subsection{Algorithm}
\label{sec:algorithm}

The algorithm used to allow the system to evolve is presented in Algorithm~\ref{generate_network}. This adopts the discrete scheme of either mimicking a collision to effect a reaction, sampling some molecules to remove from the system or adding some more molecules to it at an iteration in the simulation. 
Effecting a reaction in the system is divided into three successive steps: choosing colliding molecular pair, choosing a reaction template, and subsequently, choosing a reaction instance from the enumerated reactions.
This leads to progressive exploration of the underlying molecular space described by the specified graph grammar (i.e., set of reaction templates and set of initial molecular graphs) as the simulation progresses.
The steps in gray in Algorithm~\ref{generate_network} are optional and concern the in- and outflow parts. They can be enabled if inflows and outflows are desired, or can be skipped altogether to simulate a closed system.

\begin{algorithm}[t]
    \caption{}
    \begin{algorithmic}[1]
        \State Input: 
        \State \hspace{1cm} initial state $s_0$
        \State \hspace{1cm} number of simulation steps $N$
        \State \hspace{1cm}  \texttt{get\_rxns} method returning reactions by applying a template to a pair of molecules
        \State \hspace{1cm} inflow vector $\delta$
        \State \hspace{1cm} stochastic rate constants $k_0$, $k_1$ and $k_2$
        \State $i \leftarrow 0$ 
        \State $s[i] \leftarrow s_0$
        \While{ $i \leq N$}
            \State i++
            \State \textcolor{gray}{calculate $\mathcal{P}_0$, $\mathcal{P}_1$ and $\mathcal{P}_2$ using \eqref{zero_order}, \eqref{first_order} and \eqref{second_order} \Comment{relative probabilities}}
            \State \textcolor{gray}{choose a random number $\rho \in [0,1)$}
            \If{\textcolor{gray}{$\rho < \mathcal{P}_0/\sum_{i=0}^2 \mathcal{P}_i$}} \textcolor{gray}{\Comment{inflow block}}
                \State \textcolor{gray}{$s[i+1] \leftarrow s[i] + \delta$ \Comment{add inflow}}
        	\ElsIf{\textcolor{gray}{$\rho < \left[\mathcal{P}_0 + \mathcal{P}_1\right]/\sum_{i=0}^2 \mathcal{P}_i$}} \textcolor{gray}{\Comment{outflow block}}
                \State \textcolor{gray}{choose $m$ from $s[i]$  \Comment{outflow candidate}}
        		\State \textcolor{gray}{assemble outflow $\epsilon = [m]$}
                \State \textcolor{gray}{$s[i+1] \leftarrow s[i] - \epsilon$} \textcolor{gray}{\Comment{subtract outflow}}
            \Else \Comment{collision-reaction block}
		    	\State choose $(m_1, m_2)$ from $s[i]$  \Comment{collision pair}
	        	\State choose $r$ from $R$ \Comment{reaction template}
	        	\State \texttt{rxns} = \texttt{get\_rxns}$(m_1, m_2, r)$\Comment{return possible reactions} 
		        \If{\texttt{rxns}.length $> 0$}
		    	    \State Choose a \texttt{rxn} (defining $\nu^-$ and $\nu^+$) out of \texttt{rxns} randomly
		    	    \State $s[i+1] \leftarrow s[i] - \nu^- + \nu^+$ \Comment{update the current state} 
		        \EndIf
            \EndIf
        \EndWhile
        \State return $s$
    \end{algorithmic}
    \label{generate_network}
\end{algorithm}

The selection of $m_1$ and $m_2$ in line~15 is done according to Equation~\eqref{react_prob_different_molecules} and ~\eqref{react_prob_same_molecules}. The selection of a reaction $j$ to execute is done in line~17 by the method \texttt{get\_rxns(),} which implements the two step process described in Section~\ref{reaction_enumerate}.

Note that the loop iterates over the collision numbers, and in each iteration an event (inflow, outflow, or a collision) is supposed to occur.
However, the collisions between the molecules do not occur at roughly equal intervals of time,
if the total number of molecules in the system changes and the simulations runs under constant volume conditions.
Considering the system to be an ideal gas (where all molecules are of the same size, and the total volume of the molecules is negligible compared to that of the gas),
the collision frequency depends on the number of molecular pairs in the system.
This disregards all collisions involving three of more molecules simultaneously (which is anyway rare).
Therefore, to keep track of the time, the time step has to be weighted by the number of molecular pairs in the system.
If the system started off with $N_0$ molecules initially, and has $N$ molecules at time $t$, then the increment in real time would be
\begin{equation*}
    t \leftarrow t + \Delta t \cdot \frac{N_0(N_0-1)}{N(N-1)}
\end{equation*}
or the time interval between two collision events scales inversely with the number of molecular pairs in the system. However, if the system is simulated under constant pressure conditions (as is the case in most natural situations), the volume of the system adjusts, keeping the collision frequency roughly unchanged.

\subsection{Caching Scheme}
\label{cacheScheme}

Stochastic simulations in molecular networks naturally exhibit temporal locality in the reactions they explore. At any given stage of the simulation, the system state is typically dominated by a small subset of molecular classes with high counts, while the rest of the molecular classes have a low abundance. As a result, collisions repeatedly involve the same molecular pairs, and the same reaction templates are applied to these pairs across many consecutive simulation steps. This effect is particularly pronounced during early and intermediate phases of the simulation, where a limited set of precursor molecules drives most of the reactivity, as well as in regimes where the system approaches steady state. Consequently, the expensive operation of enumerating reaction instances for a given molecular pair and reaction template is often redundantly repeated, even though the underlying graph-theoretic computation yields identical results each time. Exploiting this temporal locality through caching allows the simulation to reuse previously computed reaction instances, thereby accelerating exploration of large molecular spaces without altering the stochastic trajectory of the system.

In more detail, the framework employs a caching strategy in which reaction instances generated by applying a template $r$ to a molecular pair $(m_1, m_2)$ are cached in a dictionary keyed by the ordered tuple $(m_1, m_2, r)$. Subsequent requests involving the same pair and template are served directly from the cache, bypassing recomputation.  This avoids repeated subgraph isomorphism checks between the left-hand side of the reaction template and the union of the molecular graphs $m_1 \cup m_2$, as well as the subsequent enumeration of product graphs resulting from valid instantiations of the template. 
Since subgraph matching and transformation of the union of the molecular graphs $m_1 \cup m_2$ to the product graphs are among the most computationally expensive operations in rule-based chemical simulations, \emph{memoizing} their outcomes can significantly accelerate the simulation, particularly in large systems where these operations are invoked frequently. This form of memoization is conceptually distinct from earlier caching approaches used in kinetic Monte Carlo simulations, which typically store interaction energies between particles and lattice sites~\cite{kmc_cache}. In contrast, the present framework caches reaction structure, enabling acceleration of stochastic simulations in generative, graph-based chemical spaces.

A \emph{cache hit} occurs when the reaction instances for a requested molecular pair and template are found in memory; otherwise, the request results in a \emph{cache miss} and the reaction instances are recomputed. Cache lookups are substantially faster than recomputation, provided that the cache data structures are designed to support efficient indexing and retrieval. The specific data structure and retrieval policy used in the cache are described in Appendix~\ref{crp}.

As we demonstrate empirically in Section~\ref{cache_effects}, reuse of cached reaction instances does enable measurable speed-ups in the simulation.
However, increasing the cache size does not proportionally improve performance. The speed-up becomes increasingly marginal as the cache size is increased: larger caches incur higher lookup and management costs, which can partially offset the gains from avoiding recomputation.
Moreover, once the cache reaches its maximum capacity, newly generated reaction instance blocks must overwrite existing entries. In this work, a least-recently-used (LRU) replacement policy is employed, in which the reaction instance block that has not been accessed for the longest time is replaced. More sophisticated policies could incorporate access frequency or attempt to predict future reuse. Ideally, reaction instances that are unlikely to be requested again would be evicted preferentially, but such foresight is difficult to achieve in a stochastic and evolving chemical system.

An additional positive consequence of using a caching scheme is that it stores a partial view of the molecular network explored during the simulation. Although the simulation algorithm in Algorithm~\ref{generate_network} does not require explicit construction or storage of the full underlying network, the cache implicitly retains a subnetwork consisting of frequently sampled reaction instances. By varying the cache size and replacement policy, different regions of the explored molecular space can be preferentially retained, without altering the stochastic trajectory of the simulation itself. This decoupling of network storage from system dynamics provides a mechanism for selectively recording and analysing portions of large molecular reaction networks while preserving the generative nature of the simulation.

\section{Experiments}
\label{exp}

In this section, we use the formose reaction as a case study to assess the behaviour of the exploration framework on a chemically rich generative system. The purpose of these experiments is not to validate quantitative reaction kinetics, since the method does not aim to reproduce them, but to examine which regions of the reachable molecular space are discovered under different exploration conditions and how open-system dynamics affect that discovery process. We therefore analyse observables such as molecular-size distributions, diversity of molecular classes, innovation of newly encountered molecules, and the abundances of selected biologically relevant products~\cite{formose_prebiotic}. These observables characterize the exploratory trajectory induced by the sampling policy and illustrate the kinds of chemically interpretable questions the framework can address.

Simulations to study the formose reaction network were carried out under two conditions: a closed system and a system with inflows and outflows. In the flow-enabled case, the fixed inflow vector $\delta$ consisted of methanal only.
We initialized the system with $990$ molecules of methanal and $10$ molecules of 2-hydroxyethanal ($1 \%$ mol) when flows were disabled. When flows were allowed, $10$ molecules of 2-hydroxyethanal were found to not have the sufficient catalytic activity for the system to evolve, as all these molecules were often removed by outflow steps even before they could react with methanal. Therefore, we initialized the system with $950$ molecules of methanal and $50$ molecules of 2-hydroxyethanal ($5 \%$ mol), that is, with a higher number of catalyst molecules, when flows were allowed. The reaction templates consisted of the reaction templates for keto–enol tautomerism, aldol addition, and their reverse reactions, shown in the four subfigures of Figure~\ref{fig:reactionTemplate}. Graph representations of molecules do not encode stereochemistry; therefore, all aldopentose stereoisomers (arabinose, lyxose, ribose, and xylose) are represented by isomorphic graphs. In both cases, the simulation consisted of $10^6$ iteration steps, and equilibrium was operationally defined as the point at which each molecular count became stable.

\begin{figure}
    \centering
    \begin{subfigure}{0.45\textwidth}
        \centering
        \includegraphics[width=\textwidth]{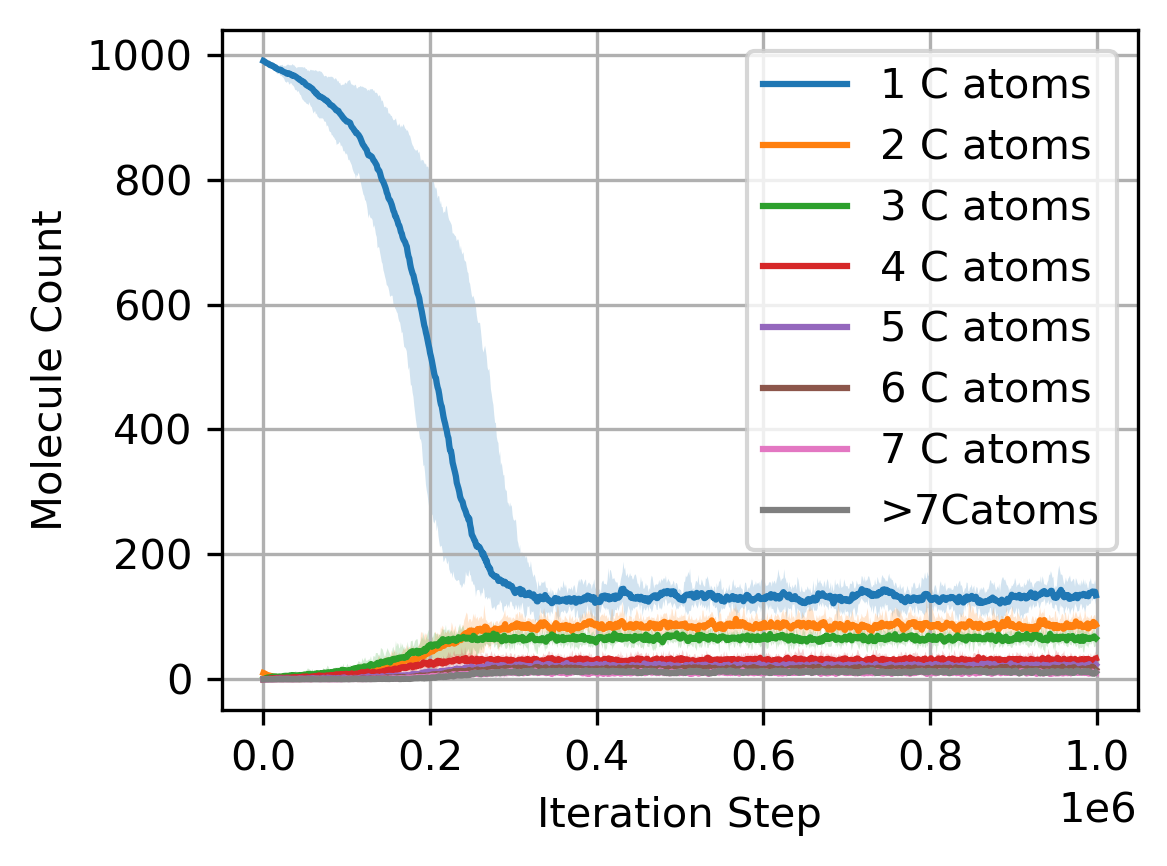}
        \caption{In the closed system, the molecules form larger molecules from methanal. See also Figure~\ref{fig:distMolsNoFlowZoom} for a zoomed-in view.}
        \label{fig:distMolsNoFlow}
    \end{subfigure}%
    \hfill
    \begin{subfigure}{0.45\textwidth}
        \centering
        \includegraphics[width=\textwidth]{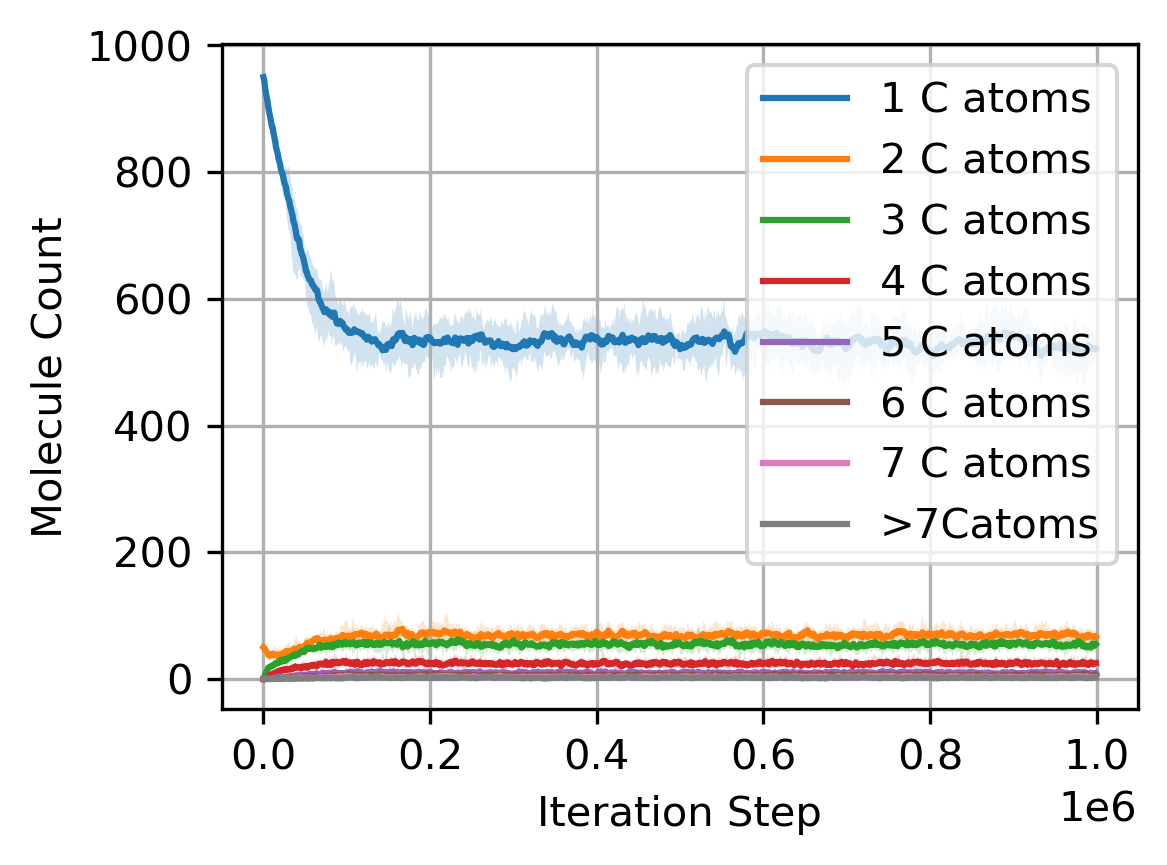}
        \caption{In the flow-enabled system, smaller molecules are more abundant. See also Figure~\ref{fig:distMolsFlowZoom} for a zoomed-in view.}
        \label{fig:distMolsFlow}
    \end{subfigure}
    \\
    \vspace{0.5cm}
    \begin{subfigure}{0.45\textwidth}
        \centering
        \includegraphics[scale=0.65]{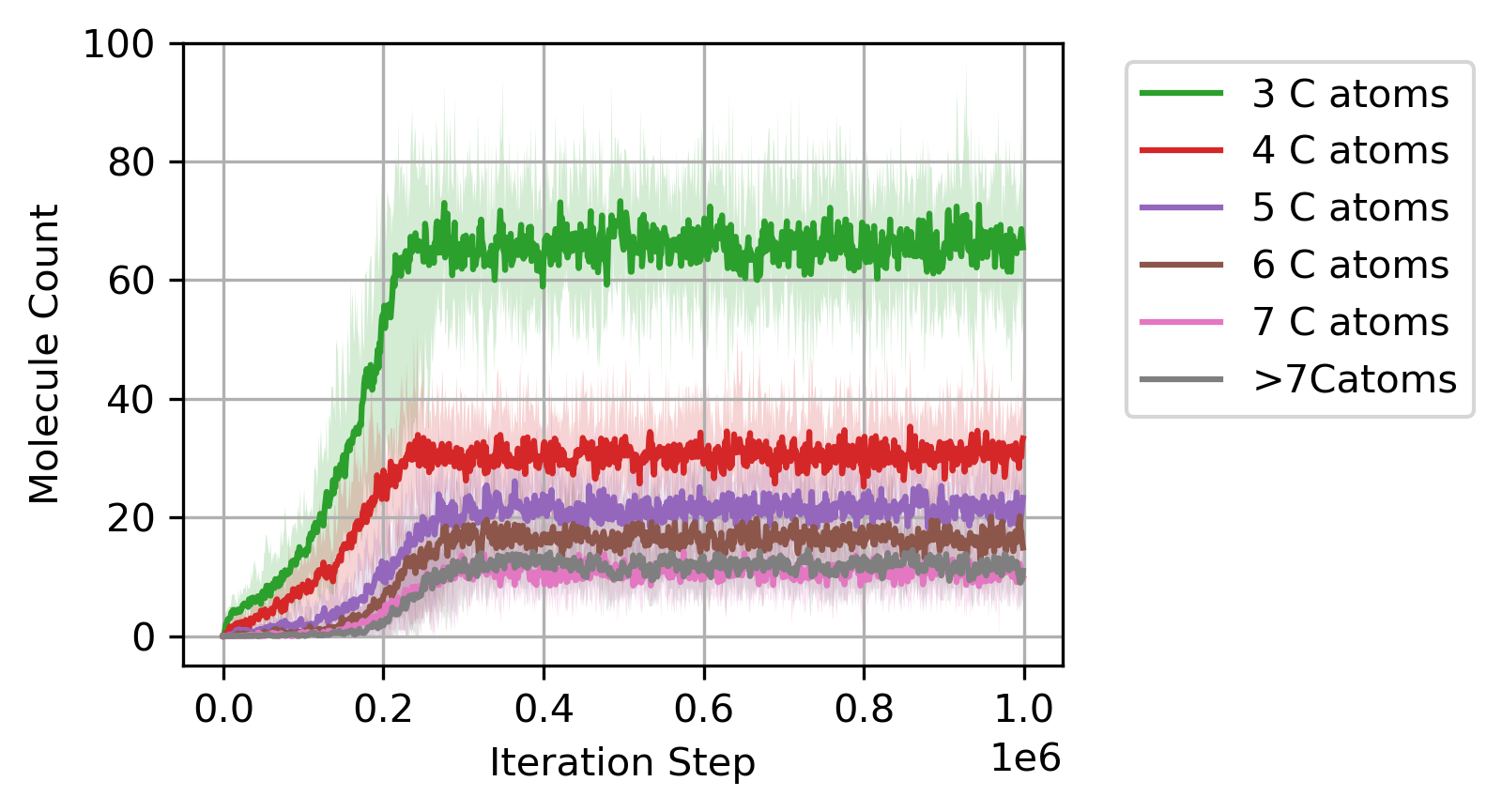}
        \caption{A zoomed-in part of Figure~\ref{fig:distMolsNoFlow} showing the molecules with five, six or more carbon atoms are more abundant than Figure~\ref{fig:distMolsFlowZoom}.}
        \label{fig:distMolsNoFlowZoom}
    \end{subfigure}%
    \hfill
    \begin{subfigure}{0.45\textwidth}
        \centering
        \includegraphics[scale=0.65]{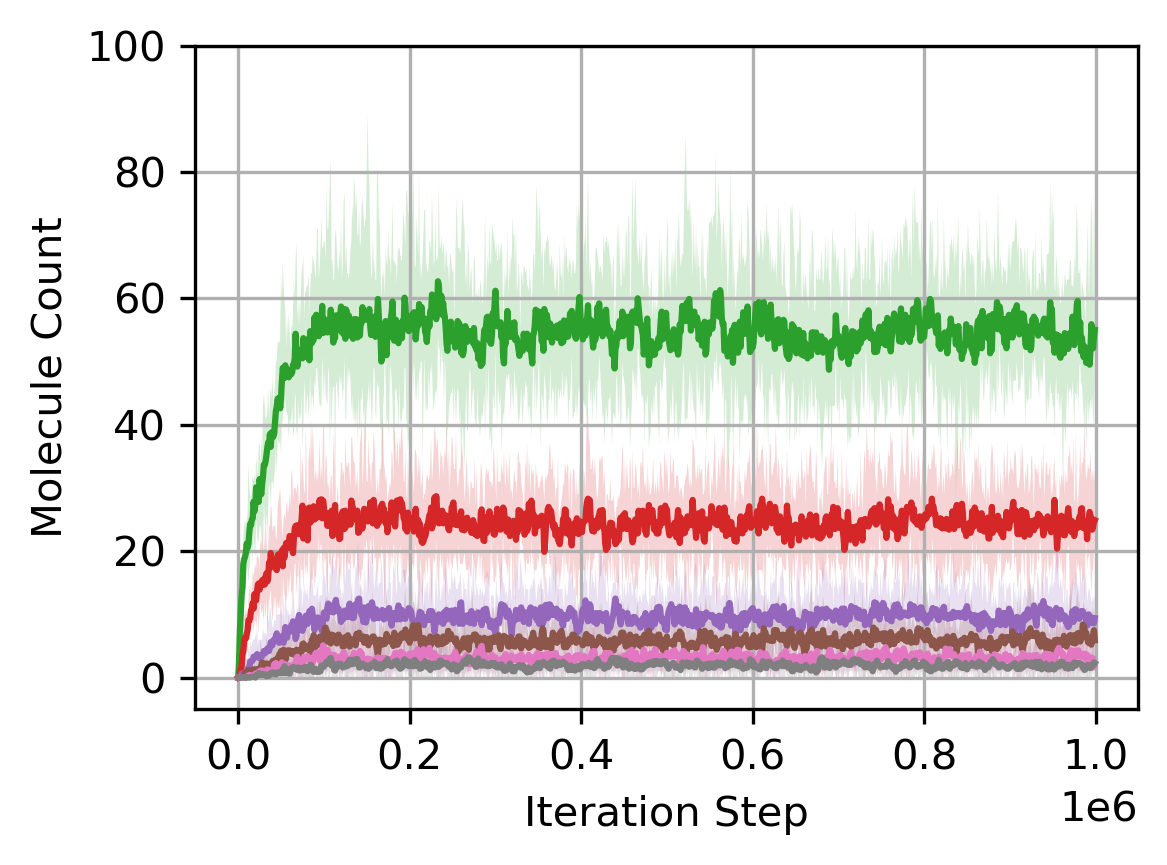}
        \caption{A zoomed-in part of Figure~\ref{fig:distMolsFlow} showing the molecules with five, six or more carbon atoms are less abundant than Figure~\ref{fig:distMolsNoFlowZoom}.}
        \label{fig:distMolsFlowZoom}
    \end{subfigure}
    \caption{The plots show the evolution of the numbers of molecules of different sizes during the simulation. If there are multiple classes of molecules with the same number of carbon atoms, they are summed up and the mean total count of molecules with a particular number of carbon atoms is plotted. All molecules with more than $7$ carbon atoms are lumped in a single bin. The values plotted are averages over $100$ trials and the shaded region around a solid curve shows the standard deviation.}
    \label{fig:distMols}
\end{figure}
Figure~\ref{fig:distMols} summarizes the evolution of molecule counts by carbon number for each scenario, with the closed system shown in Figure~\ref{fig:distMolsNoFlow} and Figure~\ref{fig:distMolsNoFlowZoom} and the flow-enabled system in Figure~\ref{fig:distMolsFlow} and Figure~\ref{fig:distMolsFlowZoom}.

In the closed system, the methanal population undergoes a sharp decline after an initial lag period of $\sim 0.5\times 10^5$ steps. The count eventually stabilizes at approximately $\sim 130\pm 10$ after $\sim 4\times 10^5$ steps. During the early phase of methanal depletion, molecules with three carbon atoms accumulate. This continues until the first $\sim 3\times 10^5$ steps, as the initial 2-hydroxyethanal population is rapidly incorporated into larger species—forming three-carbon products via aldol addition with methanal, and four-carbon products by aldol reactions with its enol tautomer (ethene-1,2-diol). As a result, three- and four-carbon molecules outnumber two-carbon molecules initially in the simulation.

\begin{figure}
    \centering
    \begin{subfigure}{0.45\textwidth}
        \centering
        \includegraphics[width=\textwidth]{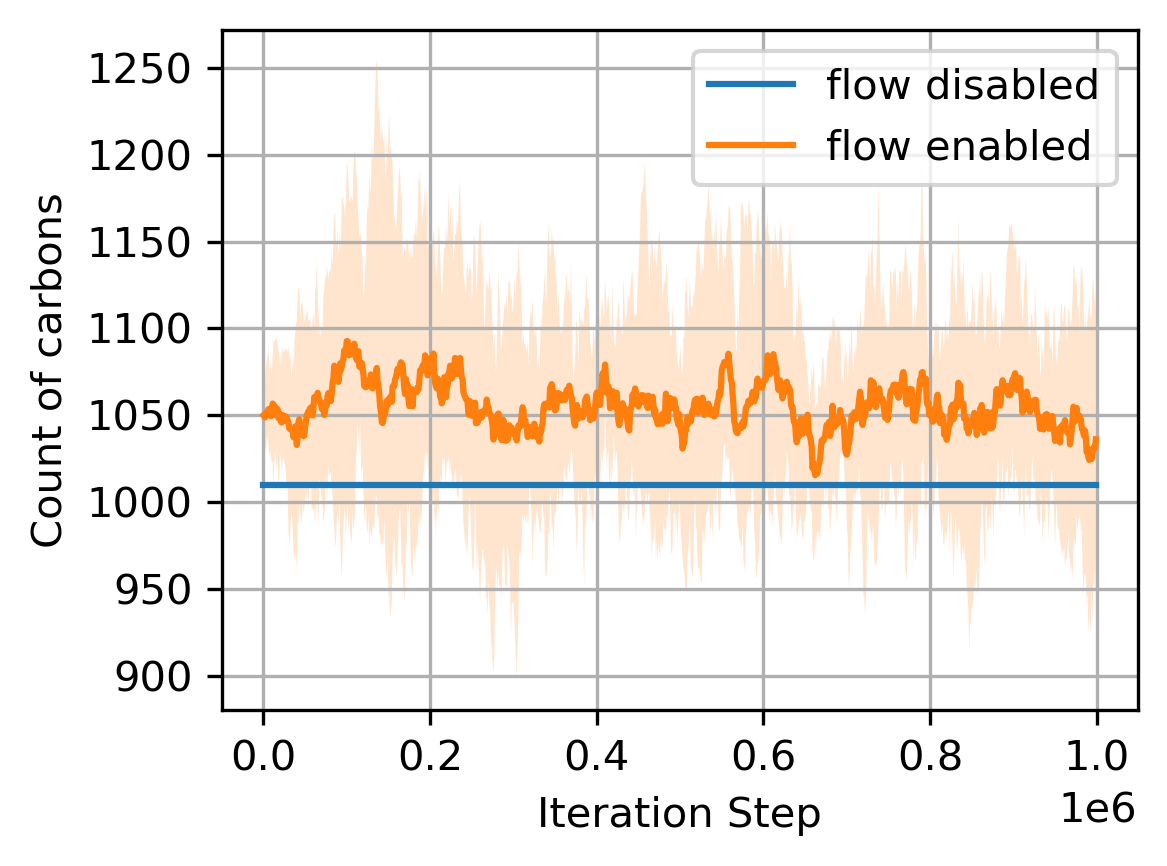}
        \caption{Number of carbons in the system.}
        \label{fig:massMols}
    \end{subfigure}%
    \hfill
    \begin{subfigure}{0.45\textwidth}
        \centering
        \includegraphics[width=\textwidth]{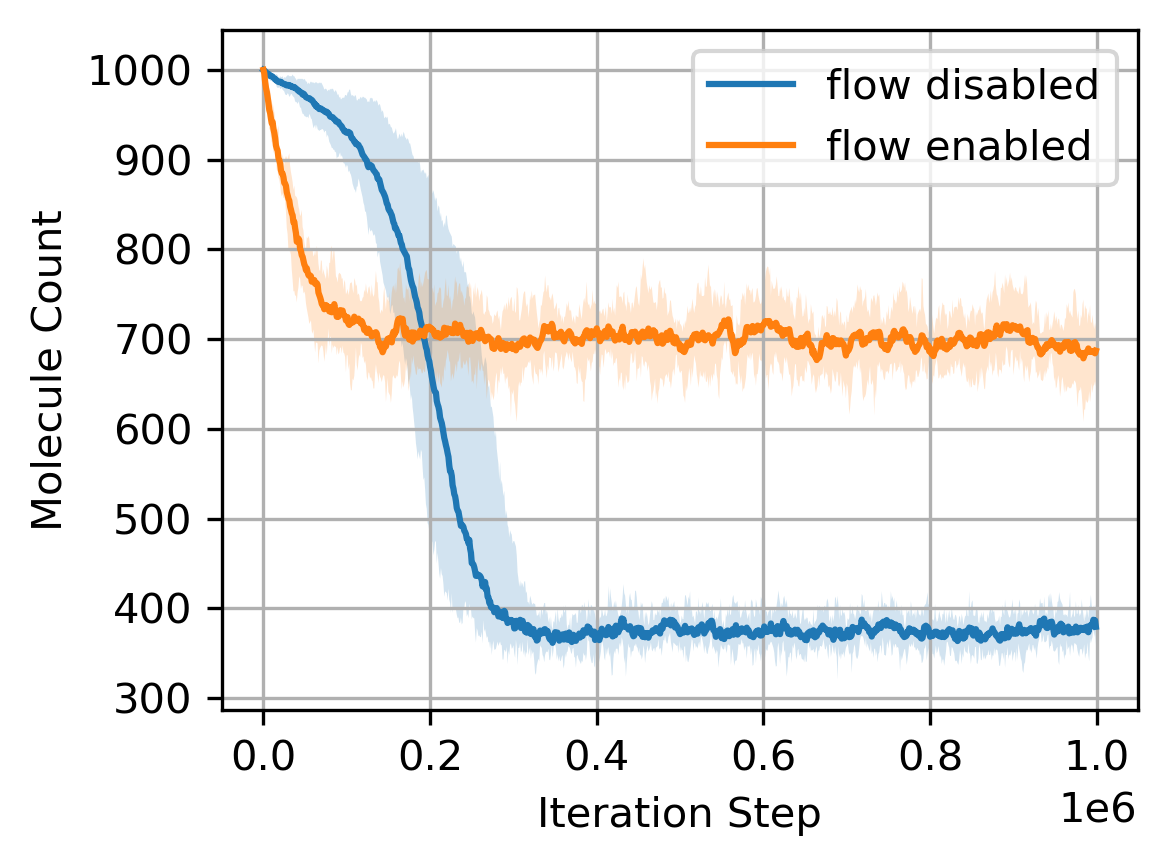}
        \caption{Number of molecules in the system.}
        \label{fig:numMols}
    \end{subfigure}
    \\
    \vspace{0.5cm}
    \begin{subfigure}{0.45\textwidth}
        \centering
        \includegraphics[width=\textwidth]{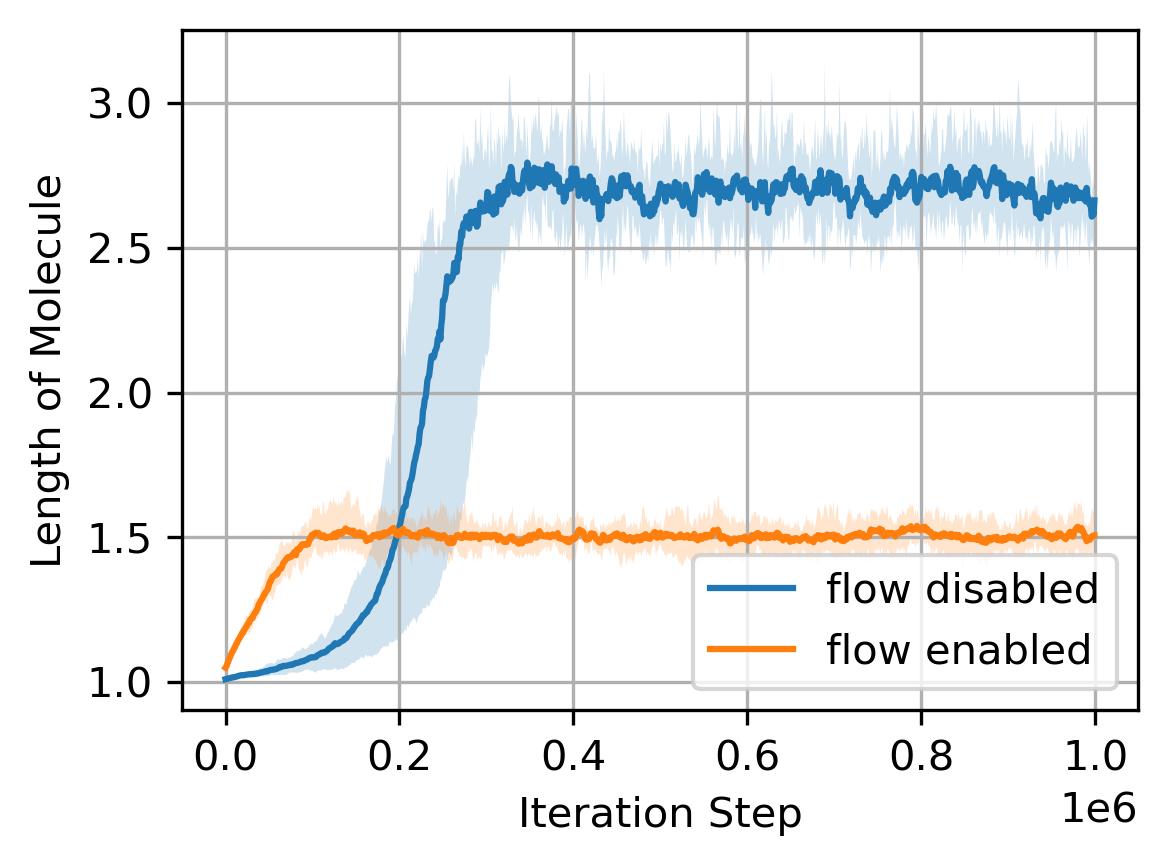}
        \caption{Average length of a molecule in the system.}
        \label{fig:lenMols}
    \end{subfigure}%
    \hfill
    \begin{subfigure}{0.45\textwidth}
        \centering
        \includegraphics[width=\textwidth]{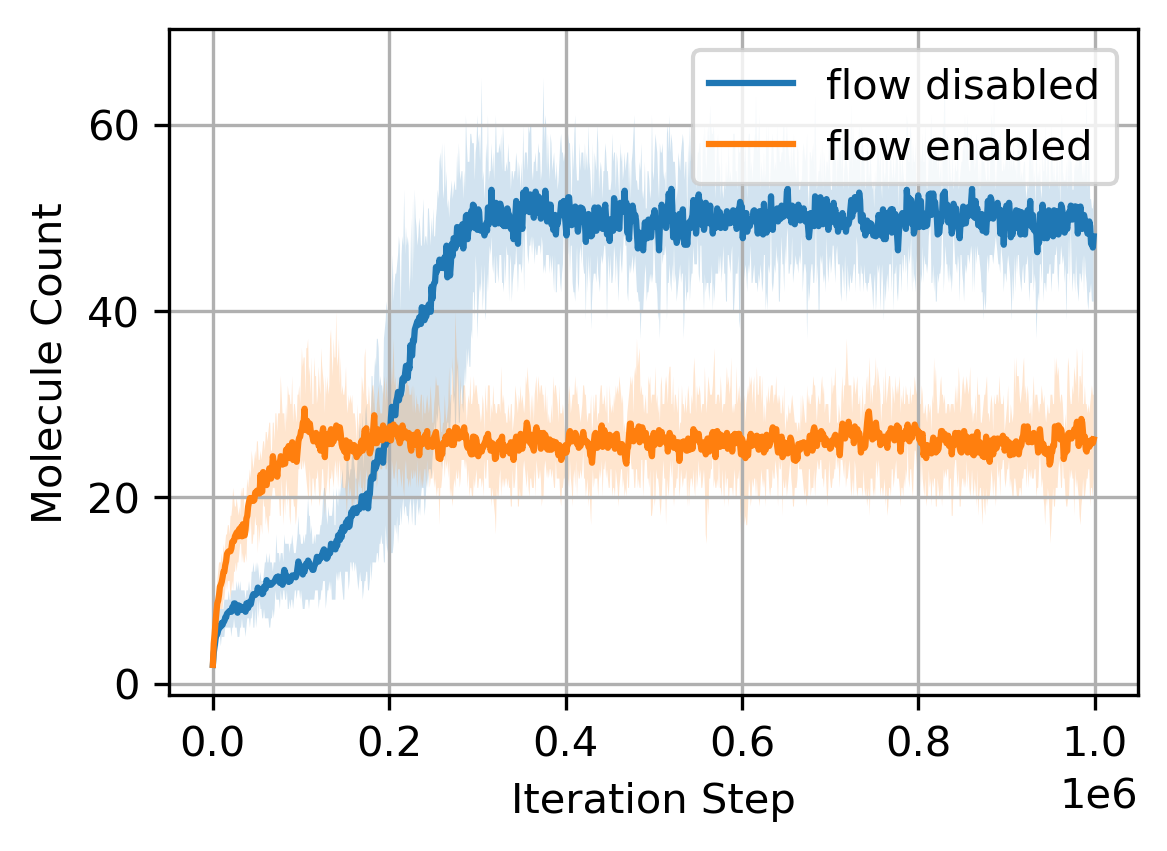}
        \caption{Number of distinct classes of molecules.}
        \label{fig:uniqueMols}
    \end{subfigure}
    \\
    \vspace{0.5cm}
    \begin{subfigure}{0.45\textwidth}
        \centering
        \includegraphics[width=\textwidth]{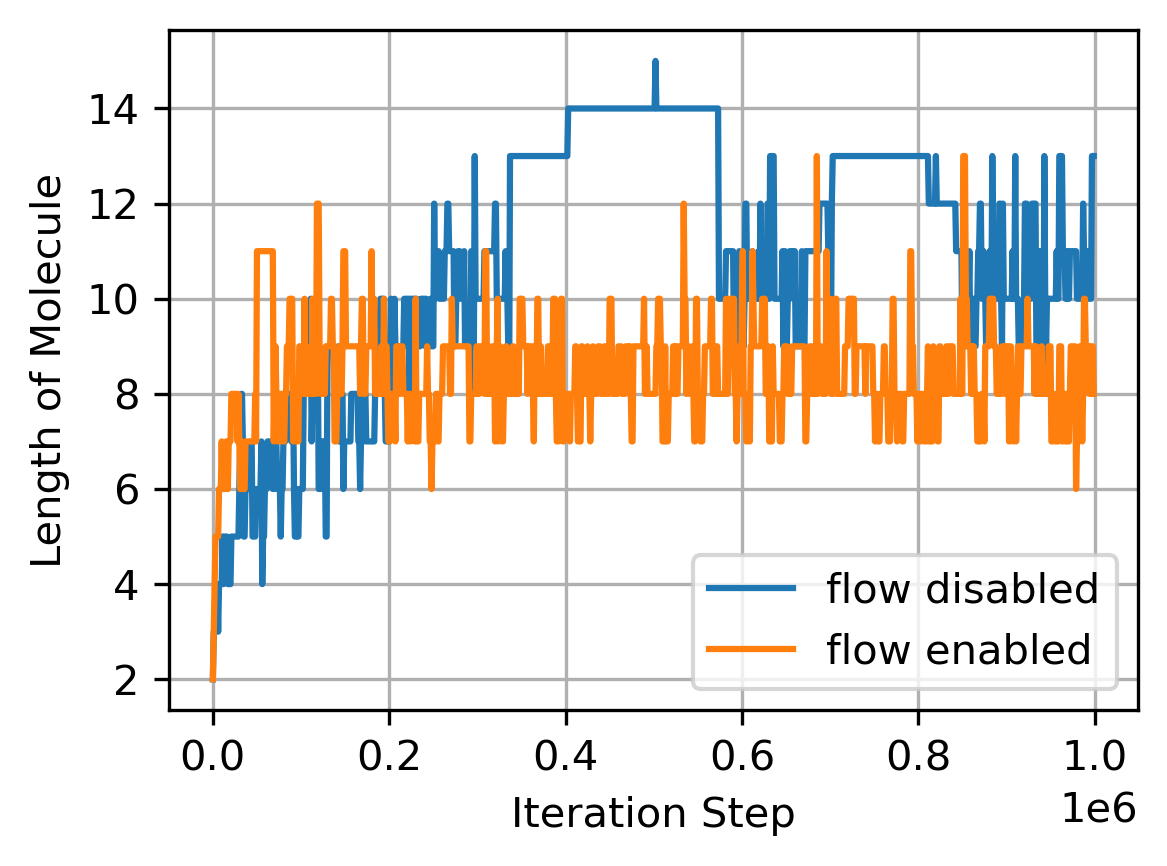}
        \caption{Length of the longest molecule in the system.}
        \label{fig:maxLenMols}
    \end{subfigure}%
    \hfill
    \begin{subfigure}{0.45\textwidth}
        \centering
        \includegraphics[width=\textwidth]{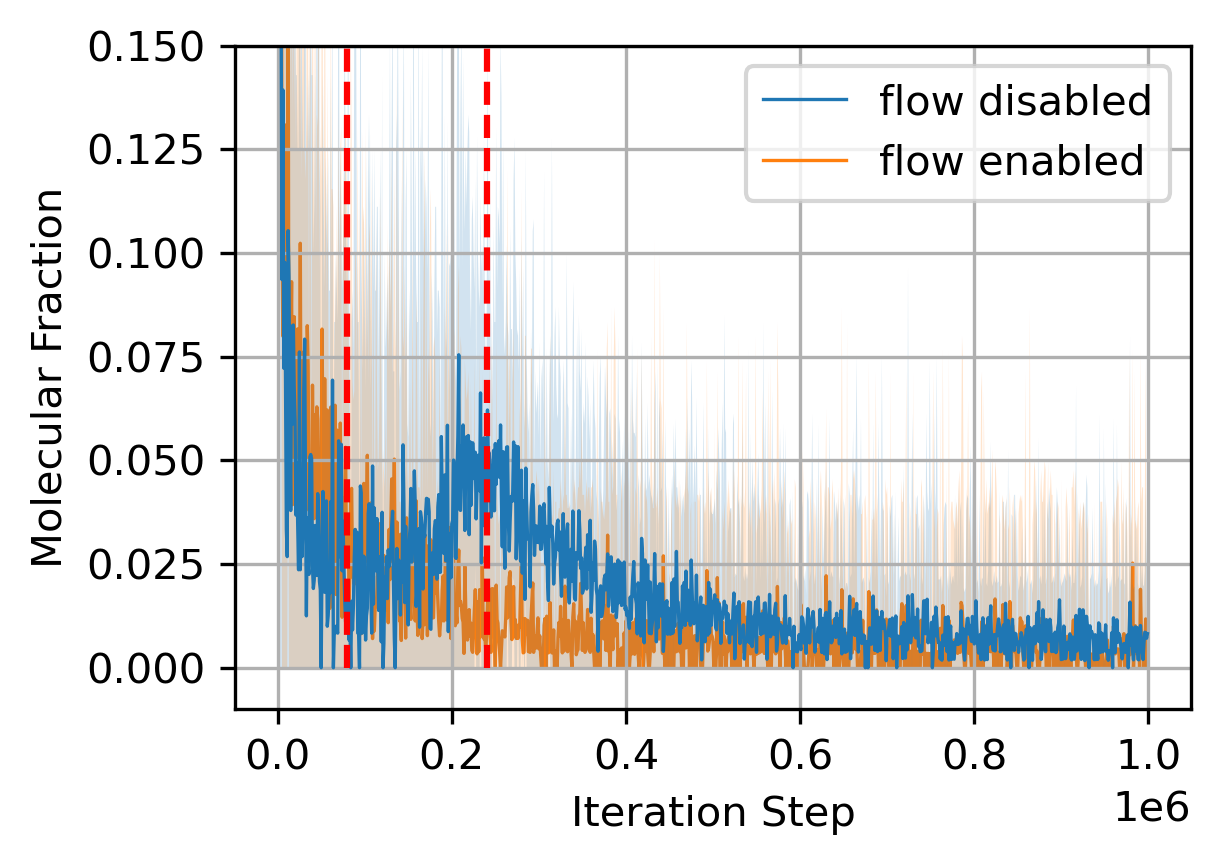}
        \caption{Fraction of new molecules.}
        \label{fig:newMols}
    \end{subfigure}
    \caption{Some parameters describing the system when the system is closed and when there is flow into and out of the system. The solid lines show the mean values and the shaded regions show the standard deviation, both over $100$ runs.}
    \label{fig:propMols}
\end{figure}

As larger intermediates fragment, however, they regenerate 2-hydroxyethanal and its tautomer (ethene-1,2-diol), increasing the abundance of two-carbon molecules in the later phase. This population ultimately stabilizes at approximately $\sim 90\pm 10$. Once the molecular distributions by carbon number have stabilized, the system exhibits no further structural evolution (as also confirmed in Figure~\ref{fig:propMols}). At this stage, the autocatalytic production of 2-hydroxyethanal becomes self-limiting: its growth ceases as soon as the ratio of available methanal to two-carbon species drops below the threshold required for autocatalysis to persist.

\begin{figure}
    \centering
    \includegraphics[width=0.5\linewidth]{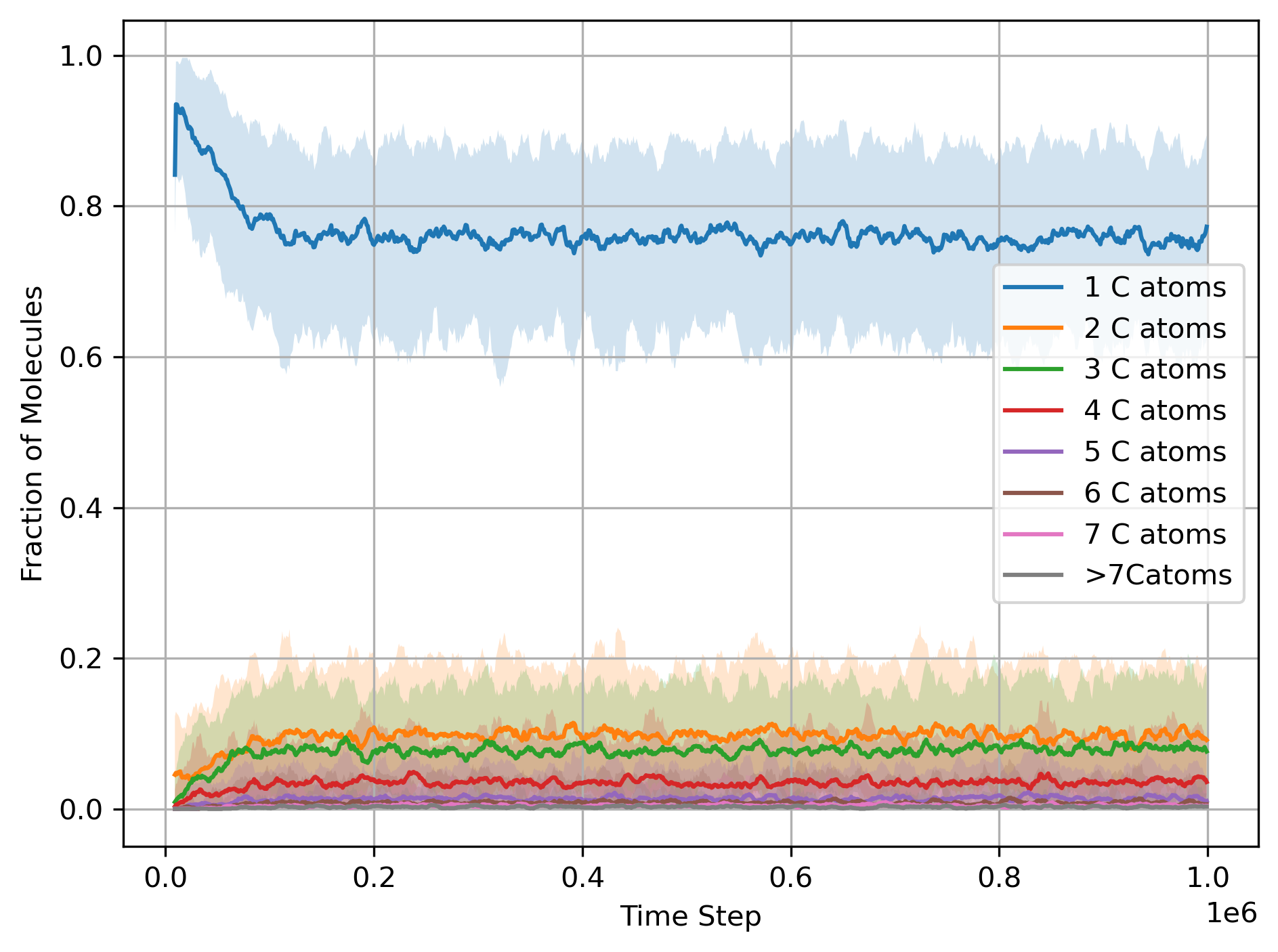}
    \caption{Fraction of molecules with different number of carbon atoms in the outflows from the system (when flows are enabled) as the simulation progresses. The values plotted are averages over $100$ trials and the shaded region around a solid curve shows the standard deviation.}
    \label{fig:outFlowMolsFrac}
\end{figure}

In the flow-enabled system, where mass exchange with the surroundings takes place, the dynamics change substantially. Methanal decreases sharply in the beginning of the simulation and then stabilizes to around $530\pm 20$ molecules after around $2\times 10^5$ simulation steps. Initially, methanal molecules are present in large numbers in the system and outflow events remove it. As methanal reacts to form other molecules, the fraction of methanal removed by outflows decreases too, as seen in Figure~\ref{fig:outFlowMolsFrac}. When equilibrium is reached, the number of methanal molecules produced in the system by reactions and added by inflow events is balanced by the number consumed by reactions and removed by outflow events.

The number of methanal molecules in the system with flows enabled is higher than in the closed system because inflow events continuously add methanal molecules to the system. Larger molecules (with five or more carbon atoms) remain consistently less abundant in the system when flows are allowed than in the closed system because outflow events remove molecules before they can grow and accumulate. Thus, even if our simulation methodology is developed with a focus more on explorability than on replication of accurate dynamics, it still well captures the differences expected to be observed in the above system with and without flows.

Figure~\ref{fig:outFlowMolsFrac} shows the variation in the number of carbons in the molecules that were removed from the system by outflow events. The number of carbon atoms in the molecules in Figure~\ref{fig:outFlowMolsFrac} closely mirrors the number of molecules with the respective number of carbon atoms in the system. This is because the outflow is concentration based as mentioned in Section~\ref{sec:inflowOutflow}, hence the probability that a molecule is removed by an outflow event depends on the number of its copies in the system.

The total number of carbon atoms in the closed system should of course remain constant (at the initial value of $1010$) throughout the simulation. This is also what was observed, as shown in Figure~\ref{fig:massMols}. Because all molecules generated from the reaction templates conform to the empirical formula (CH$_2$O)$_n$, the system’s mass can be inferred by multiplying the carbon count by $(12+16+2\times 1) = 30$. For the open system, inflow and outflow rates were tuned such that the mean total carbon count throughout the simulation remained close to the initial value of $950 + 2 \times 50 = 1050$. This maintained it close to that of the closed system, allowing for direct comparison without confounding effects from differences in system size. To accomplish this, the inflow stochastic rate constant was set to $k_0 = 75$, and the outflow stochastic rate constant to $k_1 = 0.1$. Across $100$ runs, the mean number of carbon atoms in the open system was $1055$, with a standard deviation of $43$, as depicted in Figure~\ref{fig:massMols}.

Because methanal is the overwhelmingly dominant species in both scenarios, the total number of molecules largely tracks the methanal count. In the closed system, the total population stabilizes at $\sim 375\pm 15$ after $4\times 10^5$ simulation steps, whereas in the open system it stabilizes around $\sim 700\pm 25$ after $2\times 10^5$ simulation steps. The number of molecules in the system is higher when flows are enabled as molecules are replenished by the inflow of methanal, and it stabilizes quicker than in the closed system. The total molecule count is lower in the closed system because carbon atoms accumulates in larger molecules without the possibility of replenishment.

The average molecular length (defined as the average number of carbon atoms in the molecule) differs markedly between the two conditions, as seen in Figure~\ref{fig:lenMols}. In the closed system, the average stabilizes at roughly three carbons (with mean $2.7$ and standard deviation $0.1$). In the closed system, methanal can readily be consumed to form larger molecules which raises the average length of the molecule. In the open system, however, the average settles at the lower value of $1.5$ with a standard deviation of less than $0.05$. This was to be expected: outflow tends to remove molecules before they can react to form longer molecules, while continual methanal inflow persistently skews the distribution toward shorter molecules.

The number of unique molecules—defined as the number of distinct molecular graphs in the system at any given time—diverges significantly across the two regimes (as shown in Figure~\ref{fig:numMols}). In the closed system, this count rises rapidly after an initial lag phase (the first $2\times 10^5$ simulation steps) and plateaus near $\sim 50\pm 4$. In the open system, the number of distinct molecules increases from the start and is higher than that in the closed system initially. However, the number stabilizes to a lower value of $\sim 26\pm 3$ after around $2\times 10^5$ simulation steps. This difference reflects the underlying population dynamics. In the closed system, new molecules typically arise from existing ones, since even molecular classes with few instances can react to form molecules from new classes. In contrast, the continuous outflow of molecules removes classes of molecules with more carbon atoms entirely from the system. This prevents them from forming larger and more complex counterparts, leading to a lower number of unique molecules in the system. This lower number of distinct molecular classes in the open system combined with the observation from Figure~\ref{fig:numMols} that the total number of molecules is higher implies that there are more copies of molecules of each class (in particular methanal) when flows are allowed.

Figure~\ref{fig:maxLenMols} traces the length of the longest molecule (defined as the number of carbon atoms in the molecular class with the maximum number of carbon atoms) over the progress of the simulation. Because this metric is highly sensitive to single fragmentation or condensation reaction events, the figure shows data from a single representative run of the simulation rather than the mean over multiple trials—averaging would obscure the abrupt jumps that characterize this quantity. Long molecules can fragment suddenly into much smaller species, or moderate-sized molecules can condense into substantially larger ones, producing large fluctuations in this measure. The conclusion that might be drawn is that when flows are enabled, the length of the longest molecule in the system is smaller compared to when the system is closed. Initially, the length of the longest molecule is higher in the system when flows are allowed, but as the closed system enters the generative phase (after the first $2\times 10^5$ simulation steps), the longest molecule in the closed system has a more carbon atoms.

Figure~\ref{fig:newMols} shows the innovation rate, defined as the fraction of molecular classes at simulation step~$i$ that were absent in all of the previous~$i-1$ steps:
\begin{equation*}
    \text{innovation rate} = \frac{\text{number of molecular classes not in previous $i-1$ steps}}{\text{number of molecular classes in step }i}
\end{equation*}
Because the molecular network is explored through a stochastic strategy, tracking newly discovered molecules provides insight into how effectively the simulation samples the accessible chemical space. If the innovation rate were to reach zero, the system would no longer generate new molecular structures, indicating that probably the reachable portion of the network had been exhausted. Although the theoretical space of accessible molecules is unbounded, in practice very large molecules form only sporadically and are prone to fragmentation. Even in natural chemistry, monosaccharides larger than seven carbons are rarely observed (ignoring polymeric carbohydrates). So as the simulation progresses, the innovation rate should decrease gradually as more of the reachable molecular space is discovered and becomes a part of the already explored network, yet the system continues to operate in a quasi-generative regime.

In the closed system, a transient rise in innovation rate occurs between $\sim 0.8\times 10^5$ and $\sim 2.4\times 10^5$ steps, coinciding with the sharp increase in the number of unique molecules (in Figure~\ref{fig:uniqueMols}), the average length of the molecule (in Figure~\ref{fig:lenMols}) and the rapid decline in the number of methanal molecules. This generative phase in the simulation of the closed system is in Figure~\ref{fig:newMols} marked by the vertical dotted red lines. After this point, the number of new molecular classes generated becomes small but non-zero, as seen by the observed gradual decay in the innovation rate.
When flows are enabled, this generative phase does not appear in the simulation, and the innovation rate monotonically decreases as the simulation progresses.

\begin{figure}
    \centering
    \includegraphics[width=0.5\linewidth]{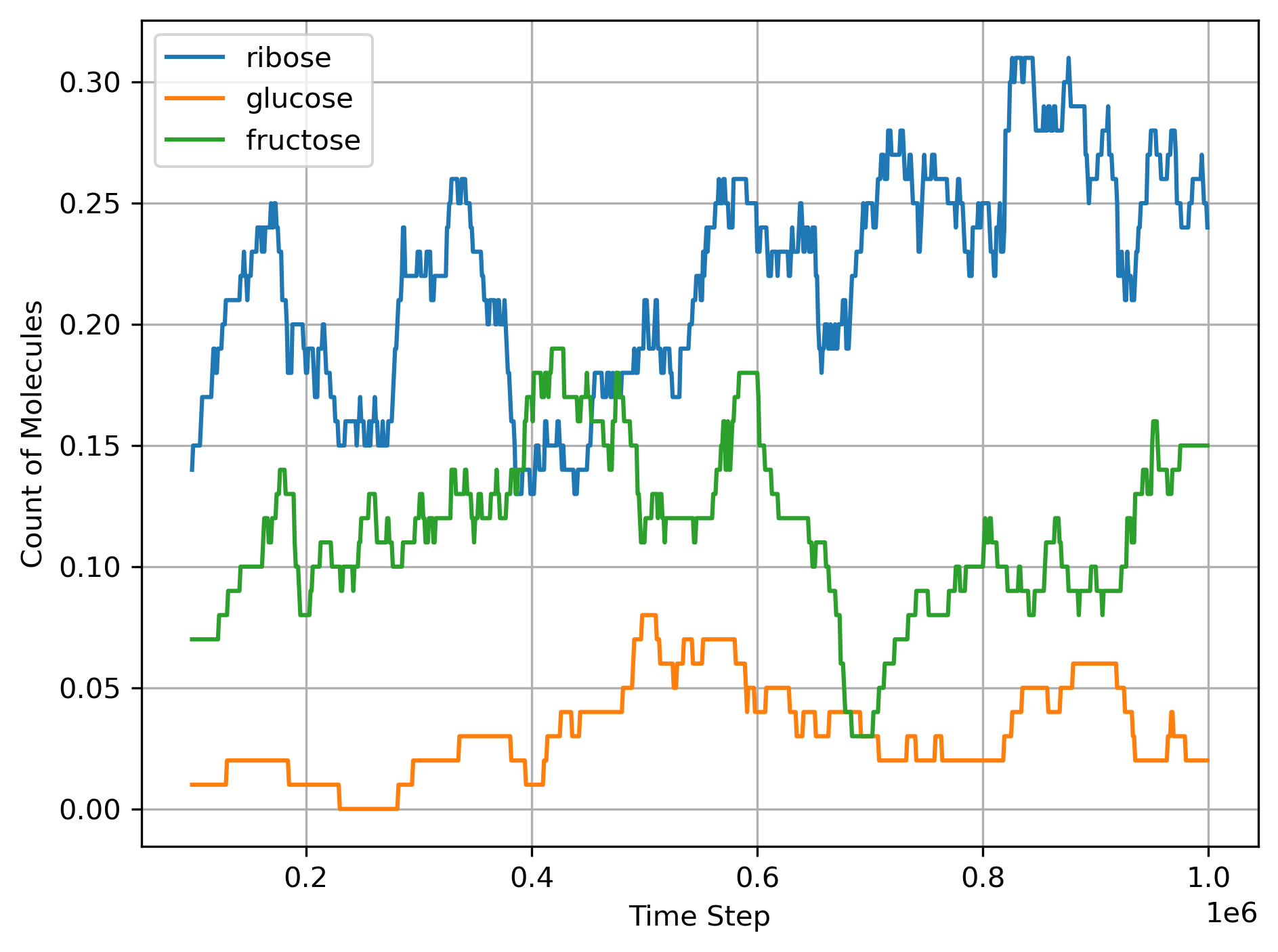}
    \caption{Counts of three key molecules in the outflow with the progress of the simulation. The plot depicts the accumulated number of molecules in the outflows every $1000$ steps averaged over $100$ trial runs of the simulation.}
    \label{fig:outflow5c6c}
\end{figure}

The formose system produces carbohydrates, including several biologically significant sugar molecules. Among the products generated in our simulations is the five-carbon sugar ribose, an essential component of ATP, NAD and RNA nucleotides. The system also produces six-carbon sugars such as glucose and fructose. Given their central biological roles---glucose as the metabolic fuel in respiration and fructose as a common plant sugar---we examine in detail how different five and six-carbon molecules arise and accumulate in the two simulation regimes.
Figure~\ref{fig:outflow5c6c} depicts the number of molecules of ribose (blue), glucose (orange) and fructose (green) recorded in the outflow from the system. The outflow count for ribose is higher than for fructose which is in turn higher than for glucose. The reasons for this observation are given later in this section.

We note that in reality, there might not be that high a number of molecules with five or six carbon atoms because larger molecules fragment fairly easily. Indeed, the ability to focus on those biologically relevant molecules highlights the explorative potential of a simulation system like the one presented here, which is not bound to necessarily follow the true dynamics of the chemical system. Algorithm~\ref{generate_network} allows us to steer the simulation precisely to this region of the network, in order to study the production of these molecules along with the reactions that might lead to them.

\begin{figure}
    \centering
    \begin{subfigure}{0.45\textwidth}
        \centering
        \includegraphics[width=\textwidth]{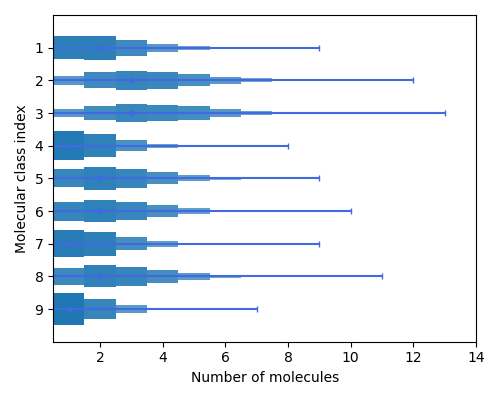}
        \caption{Distribution of number of molecules with five carbon atoms among the various molecular classes with five carbon atoms, when flows are not enabled.}
        \label{fig:5cDistNoFlow}
    \end{subfigure}%
    \hfill
    \begin{subfigure}{0.45\textwidth}
        \centering
        \includegraphics[width=\textwidth]{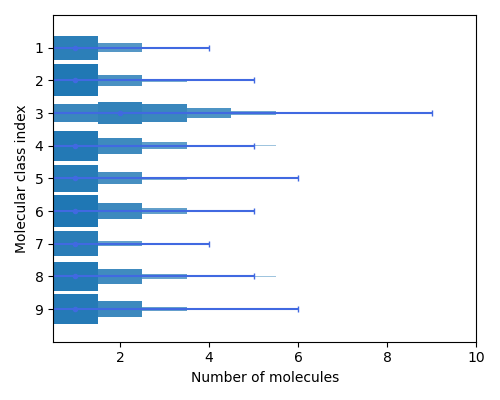}
        \caption{Distribution of number of molecules with five carbon atoms among the various molecular classes with five carbon atoms, when flows are enabled.}
        \label{fig:5cDistFlow}
    \end{subfigure}
    \\
    \vspace{0.5cm}
    \begin{subfigure}{0.45\textwidth}
        \centering
        \includegraphics[width=\textwidth]{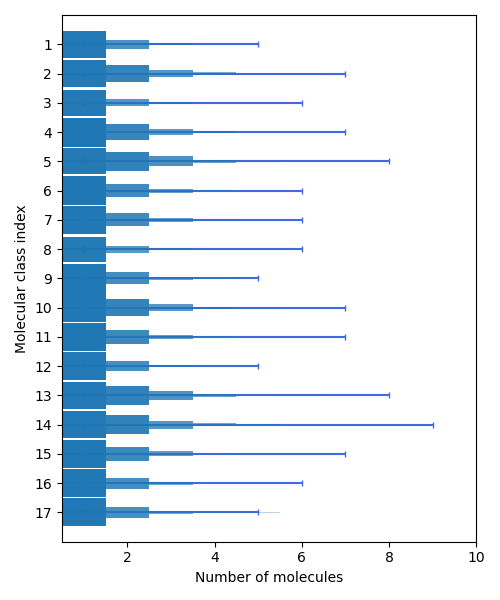}
        \caption{Distribution of number of molecules with six carbon atoms among the various molecular classes with six carbon atoms, when flows are not enabled.}
        \label{fig:6cDistNoFlow}
    \end{subfigure}%
    \hfill
    \begin{subfigure}{0.45\textwidth}
        \centering
        \includegraphics[width=\textwidth]{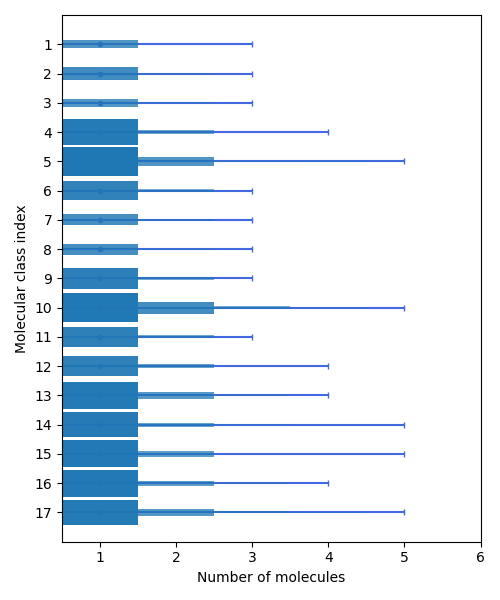}
        \caption{Distribution of number of molecules with six carbon atoms among the various molecular classes with six carbon atoms, when flows are enabled.}
        \label{fig:6cDistFlow}
    \end{subfigure}
    \caption{Distributions of counts for molecule classes with five and six carbon atoms in the closed (left) and open (right) systems. The widths of the thick lines are proportional to the number of steps in the simulation where the molecule class in question has the count value stated on the $x$-axis. The fixed-width lines extending towards the right and resembling error bars are there to visualize very small line widths, in order to show the maximum number of molecules in that class during the simulation.}
    \label{fig:histMols}
\end{figure}

Figure~\ref{fig:histMols} plots and compares the number of molecules in the different molecular classes with five carbons (top) and six carbons (bottom) for the closed system (left) and the open system (right). These plots show the distribution of the (discrete) counts of the number of molecules for each molecular class over the simulation. The structures corresponding to the integer labels on the y-axes appear in Tables~\ref{tab:5CMols} and~\ref{tab:6CMols}. For example, the molecular graph for molecular class~1 among the molecules with five carbon atoms represents two stereoisomers corresponding to ribose, while the molecular graph for class~1 among the molecules with six carbon atoms contains four stereoisomers corresponding to glucose and galactose (two enantiomeric pairs). The molecular graph for class~6 of the six-carbon molecules represents the stereochemical variants of fructose. Overall, Figure~\ref{fig:histMols} shows that the number of molecules with five or six carbon atoms in the system is less when flows are allowed. This was expected because the count of a molecular class is determined primarily by the number of reactions producing it and the availability of the molecules that form it (precursors), in the closed system. A higher number of collisions between its precursors increases the formation rate of the molecule. In the open system, by contrast, molecular classes that are formed through shorter reaction pathways (fewer transitions from the initial species) tend to be observed in higher amounts, before outflow removes them preventing them to isomerize or react to form other molecules.

\begin{figure}
    \centering
    \begin{subfigure}{\textwidth}
        \centering
        \includegraphics[scale=0.3]{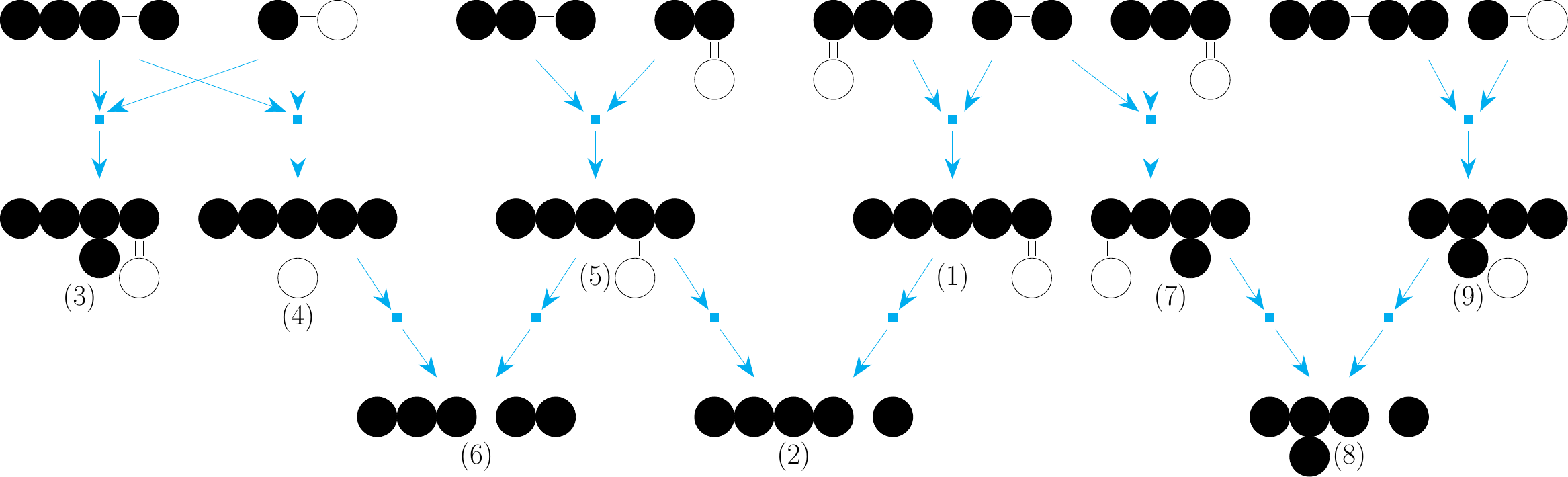}
        \caption{Reactions that form molecules with five carbon atoms are depicted in this schematic representation of a part of the system.}
        \label{fig:rn5c}
    \end{subfigure}
    \\
    \vspace{0.5cm}
    \begin{subfigure}{\textwidth}
        \centering
        \includegraphics[scale=0.3]{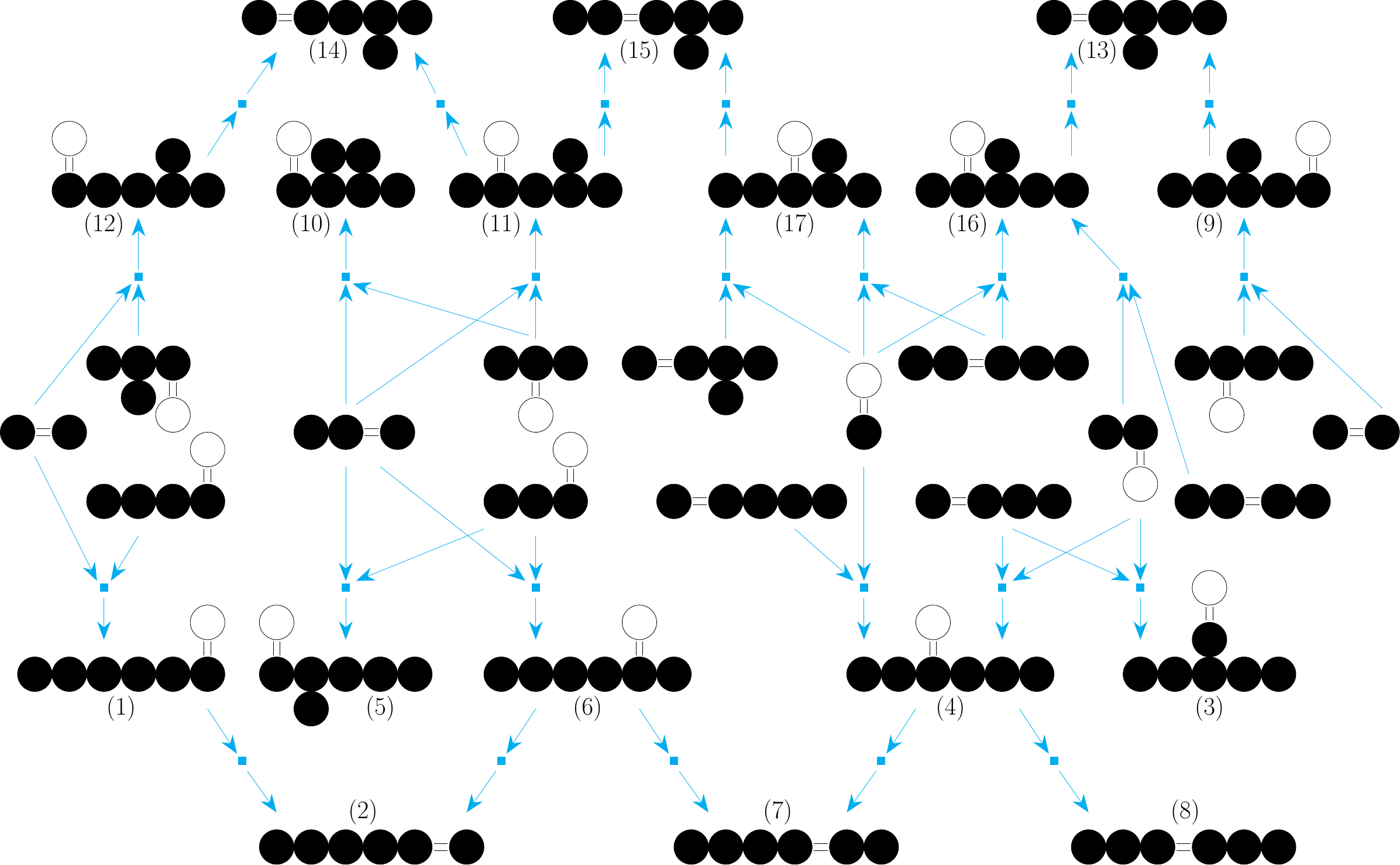}
        \caption{Reactions that form molecules with six carbon atoms are depicted in this schematic representation of a part of the system.}
        \label{fig:rn6c}
    \end{subfigure}
    \caption{Network representation of fragments of the chemical system. All the reactions are reversible but the reverse directions are not depicted to reduce clutter. The molecule can be inferred from the schematics showing carbon atoms (solid circles), double bonds, and carbonyl oxygen atoms (hollow circles).}
    \label{fig:rn56}
\end{figure}

Figure~\ref{fig:rn5c} shows the molecular network fragment involving the five-carbon species. The solid black circles depict a carbon atom, while the hollow circles depict carbonyl oxygen atoms. The hydroxyl oxygen atoms and hydrogen atoms are not depicted to reduce clutter. The position of the double bond is shown explicitly. The molecular structure can be inferred from this information as all molecules have the empirical formula (CH$_2$O)$_5$. 

In both the closed and open systems as shown in Figures~\ref{fig:5cDistNoFlow} and~\ref{fig:5cDistFlow} respectively, class~3 among the molecular classes with five carbon atoms is the most abundant. The modal count of this molecular class is $3$ and $2$, in the closed and open systems, respectively. This class, an aldehyde, does not isomerize into other five-carbon structures, so its loss is limited to fragmentation into precursor molecules, leading to relatively high steady-state levels.

\begin{table}
	\centering
	\begin{tabular}{@{}l@{\phantom{aaaaaa}}cc@{}}
	\toprule
	& Enol class & Precursor carbonyl class \\
	\midrule
	\multirow{3}{6em}{Classes with five C atoms} & 2 & 1, 5\\
	& 6 & 4, 5\\
	& 8 & 7, 9\\
	\midrule
	\multirow{6}{6em}{Classes with six C atoms} & 2 & 1, 6\\
	& 7 & 4, 6\\
	& 8 & 4\phantom{, 4}  \\
	& 13 & \phantom{9}9, 16\\
	& 14 & 11, 12\\
	& 15 & 11, 17\\
	\bottomrule
	\end{tabular}
	\caption{List of the enol molecular classes with five and six carbon atoms, with the respective carbonyl classes from which they are formed.}
	\label{tab:ketoEnolClasses}
\end{table}

In the closed system in Figure~\ref{fig:5cDistNoFlow}, the next most abundant classes are the enols (class~2, class~8 and class~6 in decreasing order). These classes are each formed from two precursor classes as listed in Table~\ref{tab:ketoEnolClasses}, and therefore act as a sink for both. In the flow-enabled system shown in Figure~\ref{fig:5cDistFlow}, however, classes~2, 8 and 6 become less abundant because outflow removes the precursor carbonyl molecules before they can isomerize into molecules of the enol classes. Their reduced residence time limits their formation into the enol. The carbonyl classes~5 and 9 become the next dominant classes as they are formed in a shorter number of steps from the starting molecules (methanal and 2-hydroxyethanal). Therefore, the distribution of the counts of the molecular classes for the open system starts becoming skewed towards less diversity.

A similar effect can be seen among the molecular classes containing six carbon atoms: flows skew the relative distribution of the molecular classes (as seen in Figure~\ref{fig:6cDistFlow}) away from the enols compared to when the system is closed (as seen in Figure~\ref{fig:6cDistNoFlow}). Figure~\ref{fig:rn6c} shows the network fragment for the formation and inter-conversion between all molecules with six carbon atoms. In the closed system, enol class~14 is the most abundant, followed by classes~5 and 13. Both 13 and 14 are enol classes formed from two precursor carbonyl class, while class~5 is an aldehyde that cannot isomerise into an enol.

The count of the molecular class~6 (which contains the isomers of fructose) is higher than that of class~1 (containing the isomers of glucose) indicated by the width of the corresponding bars, which explains the lower number of glucose molecules in the outflow compared to fructose molecules in Figure~\ref{fig:outflow5c6c}. 
The precursors for these molecular classes are expected to influence their abundances. Class~6 is formed from two molecules with three carbon atoms each, while class~1 is formed by a reaction between an aldehyde with four carbon atoms and an enol with two carbon atoms. Figure~\ref{fig:distMols} indicates that molecules with three carbon atoms are more abundant than those with four carbon atoms but less abundant than their two-carbon counterparts. Thus, the two required molecules with three carbon atoms might collide more frequently than a 4-carbon and a 2-carbon molecule (consistent with Equation~\eqref{react_prob_different_molecules}). This might be a possible explanation for why class~6 is more abundant than class~1. 

\section{Effects of Caching}
\label{cache_effects}

\begin{figure}
    \centering
    \includegraphics[width=0.8\linewidth]{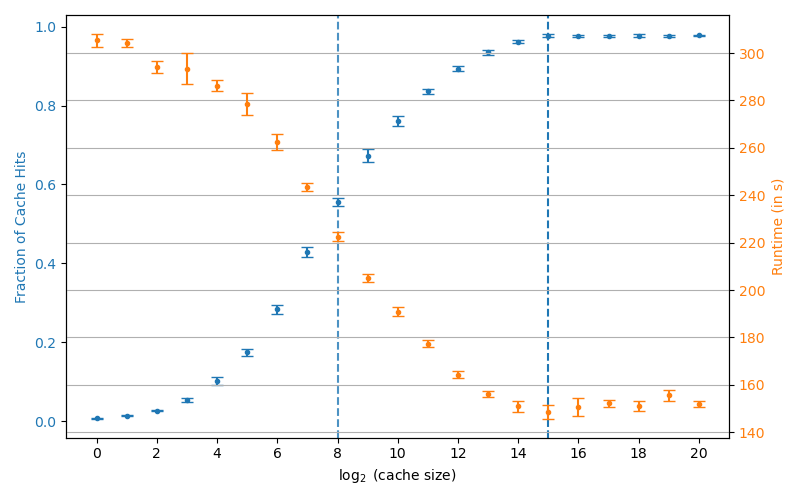}
    \caption{Effect of the cache size on the cache hit rate (blue) and the runtime of the simulation (orange). The chosen cache sizes for further experiments, $2^8 = 256$ and $2^{15} = 32768$ are denoted by the vertical dashed line.}
    \label{fig:cacheSizeEffects}
\end{figure}

To identify a cache size that optimizes performance, we on the pre-biotic formose system studied in Section~\ref{exp} performed $100$ independent simulation runs for each candidate cache size $2^i$, $i = 0, 1, 2,\dots, 20$. For each run, we recorded the wall-clock time required to execute $10^6$ simulation steps and the average cache hit rate (the fraction of requests resulting in a cache hit). Unless otherwise stated, flows are not enabled. The results are summarized in Figure~\ref{fig:cacheSizeEffects}.

As expected, increasing the cache size leads to a higher cache hit rate (blue markers). The improvement is initially rapid and reaches an inflection point at cache size of $2^8$, beyond which the improvement slows down. There is still improvement up to a cache capacity of $2^{15} = 32768$ (with the corresponding cache hit rate being $0.977\pm0.03$), after which the hit rate settles at just below the maximum possible value of one.
A similar saturation effect is observed in the simulation runtime: enlarging the cache leads to substantial speedups up from $305\pm 3$ seconds to $148\pm 3$ seconds, with an inflection point at size $2^{8}$ and no substantial improvements after size $2^{15}$.
The lack of improvement after increasing the cache size beyond $2^{15}$ probably suggests that the number of active cache keys in the system is around~$2^{15}$. On increasing the cache size beyond this, one cannot expect further improvements, and the runtime flattens out to a value dominated by the execution of the rest of the actions during the simulation. The speed-up is then determined by the relative runtime of the actions cached and the rest of the actions.

It is important to note that both the cache hit rate and the resulting speedup depend not only on cache size but also on the dynamical properties of the system being simulated and the cache replacement policy employed. Systems that rapidly explore new regions of state space and rarely revisit recently encountered molecular configurations are expected to benefit less from caching, regardless of cache size. Conversely, systems that repeatedly traverse similar local regions of the molecular space exhibit stronger temporal locality and therefore benefit more from caching. As a result, the optimal cache size can be expected to be system-dependent.

For the formose system studied here, a cache size of $2^{15}$ reaction instances was employed because further increasing the cache size did not lead to improvements in the performance. However, experiments were also performed with a cache size of $2^8$ to study the evolution of the cache hit rates during the course of the simulation, when the cache is much smaller than the size of the working set of the active cache keys in the simulation. All subsequent simulations were therefore run twice with these two cache sizes.

\subsection{Cache utilization and hit rates over time}

\begin{figure}
    \centering
    \begin{subfigure}{0.49\textwidth}
        \centering
        \includegraphics[width=\textwidth]{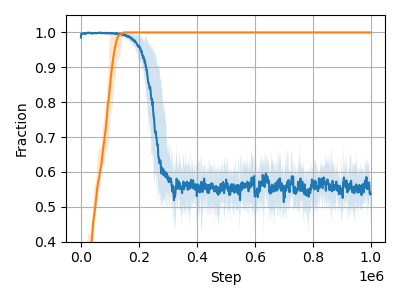}
        \caption{Flows disabled and cache size $2^8$.}
        \label{fig:cacheNoFlow}
    \end{subfigure}%
    \hfill
    \begin{subfigure}{0.49\textwidth}
        \centering
        \includegraphics[width=\textwidth]{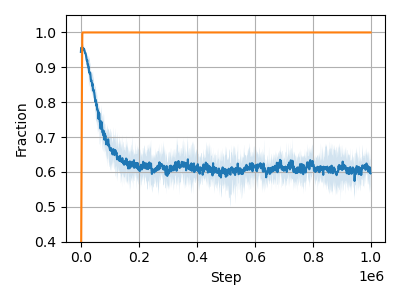}
        \caption{Flows enabled and cache size $2^8$.}
        \label{fig:cacheFlow}
    \end{subfigure}\\
    \vspace{0.2cm}
    \begin{subfigure}{0.49\textwidth}
        \centering
        \includegraphics[width=\textwidth]{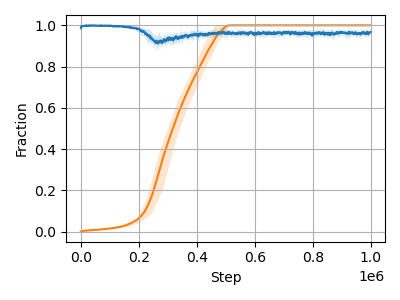}
        \caption{Flows disabled and cache size $2^{15}$.}
        \label{fig:bigCacheNoFlow}
    \end{subfigure}%
    \hfill
    \begin{subfigure}{0.49\textwidth}
        \centering
        \includegraphics[width=\textwidth]{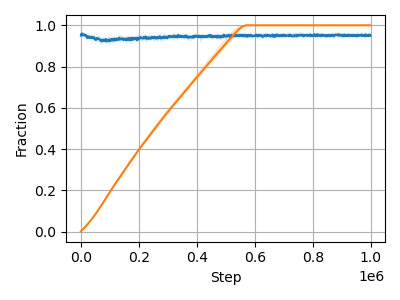}
        \caption{Flows enabled and cache size $2^{15}$.}
        \label{fig:bigCacheFlow}
    \end{subfigure}
    \caption{Cache hit rate (blue) and fraction of utilization of the cache (orange) under the two different flow conditions and two different cache sizes.}
    \label{fig:cacheAccess}
\end{figure}

With the cache size fixed at $2^8$ and $2^{15}$ reaction instances, we analysed how cache utilization and hit rates evolve during the course of a simulation. The cache hit rate (blue) and the fraction of cache capacity utilized (orange) are shown in Figure~\ref{fig:cacheAccess}. Cache hit rate and utilization are two correlated measures of cache use: once the entire cache has been utilized, new instances would be added to the cache at the expense of existing instances and the hit rate would start to fall. Prior to that, the hit rate would be close to one, because any new instances can be added to the cache which has not yet reached its capacity.

In the closed system with cache size $2^8$ (Figure~\ref{fig:cacheNoFlow}), the cache is populated gradually over the first $1\times 10^5$ simulation steps. During this initial phase, the diversity of reactions is low and the system repeatedly applies a small set of reaction templates to the small set of molecular pairs. Consequently, the cache hit rate is high (nearly $100\%$). Once the cache is filled, new reaction instances are stored by evicting the least recently used cached entries. This eviction of reaction instances leads to a sharp decline in the hit rate, as expected, which stabilizes at approximately $56\pm 12\%$ after about $3\times 10^5$ steps.
This stabilization coincides with the system ending its generative phase: by this point, the total number of molecules, the maximum molecular length, and the number of unique molecular classes have all reached steady values, as seen in Figure~\ref{fig:distMolsNoFlow}. The reduced hit rate therefore reflects increased diversity in reaction requests during the earlier generative phase, followed by repeated sampling from a larger but fairly stable set of reactions.

In contrast, when inflows and outflows are enabled with a cache of size $2^8$ (Figure~\ref{fig:cacheFlow}), the cache fills much more rapidly, within the first $2\times 10^3$ steps. The cache hit rate drops sharply during the early stages of the simulation, indicating a higher initial diversity of reactions. This decline gradually decreases and stabilizes around $61\pm 9\%$ beyond the first $2\times 10^5$ steps.
The higher steady-state hit rate in the flow-enabled system reflects the lower molecular diversity maintained in the system under open conditions, as seen in Figure~\ref{fig:distMolsFlow}. Continuous inflow of methanal and outflow of larger molecules restricts the range of molecular classes present at any given time, reducing the number of distinct molecular pairs and reaction templates encountered.

In the closed system with cache size $2^{15}$ (Figure~\ref{fig:bigCacheNoFlow}), the cache is populated over the first $5\times 10^5$ simulation steps. The cache is filled gradually in the initial phase consisting of the first $2\times 10^5$ steps, after which the cache utilization picks up pace. There is a slight dip in the cache hit rate at the beginning of this phase, where new reaction instances are explored by the system, which was not found in the cache. These are added to the cache (whose capacity is not reached yet) raising its utilization rate. The cache hit rate remains consistently high ($\sim 96\pm 3\%$) during the entire simulation, indicating that the cache at all times have capacity for holding the active set of reactions.

When flows are allowed along with using a cache size $2^{15}$ (Figure~\ref{fig:bigCacheFlow}), the cache is filled over the first $5\times 10^5$ simulation steps, with the rate being roughly similar over the entire phase. The cache hit rate is maintained at roughly $\sim 95\pm 3 \%$ over the entire simulation.

\subsection{Impact on runtime}

The primary motivation for introducing caching is to reduce the computational overhead associated with repeatedly enumerating reaction instances, particularly the costly subgraph isomorphism checks required to match reaction templates. Even though the underlying graph transformation engine MØD~\cite{mod_graphTransformation} is highly optimized, caching at the simulation level still yields measurable performance improvements. Caching does not alter the observables of the system (such as the distribution of the number of carbons in the molecules, the innovation rate for molecules or the relative abundances of different molecular classes or the trajectory of the simulation) and has no impact on the outcome of the stochastic exploration.

Table~\ref{tab:runtime} summarizes the average runtimes over 100 simulation trials, without and with caching (with different cache sizes). When flows are disabled and a cache of size $2^8$ is used, caching reduces the runtime by approximately $91$ seconds ($\sim 27\%$). With the same cache size, when flows are enabled, the reduction is approximately $59$ seconds ($\sim 16\%$).
The decrease in the runtime with cache size $2^{15}$ and without flows is approximately $187$ seconds ($\sim 56\%$), while with flows enabled, the reduction is approximately $160$ seconds ($\sim 42\%$).
Despite the higher cache hit rate observed in the flow-enabled system, the relative speedup is smaller. This is because cache hits only accelerate steps in which a chemical reaction is to be executed. Inflow and outflow steps—which constitute a substantial fraction of simulation steps when flows are enabled—do not benefit from caching. Since the cache hit rate is computed only over reaction steps, the overall fraction of accelerated steps is lower, leading to a smaller net reduction in runtime.

\begin{table}
	\centering
	\begin{tabular}{@{}lcc@{}}
	\toprule
	& Without flows & With inflows and outflows \\
	\midrule
	Caching disabled & $335\pm 7$ & $377\pm 7$  \\
	Cache size $2^8$ & $244\pm 2$ & $318\pm 5$  \\
	Cache size $2^{15}$ & $148\pm 3$ & $217\pm 2$ \\
	\bottomrule
	\end{tabular}
	\caption{Average runtime of the $100$ trials of the simulation, in seconds, without and with caching (for two different cache sizes). The results when flows are enabled are listed in the right column.}
	\label{tab:runtime}
\end{table}

\begin{figure}
    \centering
    \includegraphics[width=0.9\linewidth]{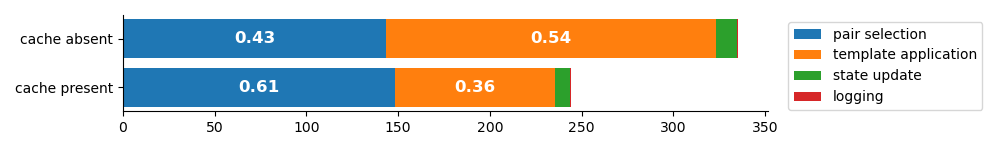}
    \caption{Breakdown of the total runtime of an average of $10$ trials of the simulation into what is spent on collision pair and reaction template selection (blue), template application or cache lookup on cache hit (orange), state update (green), and logging (red), when caching is disabled (top) and enabled (bottom). The labels (white) give the fraction of the total simulation time used for those components.}
    \label{fig:fracCacheTimes}
\end{figure}

Figure~\ref{fig:fracCacheTimes} gives the cost breakdown for the various components in the simulation, without and with a cache of size $2^{8}$. As described in Algorithm~\ref{generate_network}, choosing a transition out of the current state is divided into choosing a collision and a reaction template and then applying the chosen reaction template to the chosen molecules. Once the transition out of the current state is chosen, the state is updated and logged (if required). Without caching, the application of the reaction template on the molecular pair took more than half of the runtime. Using a cache to look up the results on a cache hit reduced template application time by a factor of two, leaving it at roughly one-third of the (now reduced) total simulation time.

With caching, we observe from Figure~\ref{fig:fracCacheTimes} that the fraction of runtime required to choose a pair of molecules from the state dictionary now is the largest part of the total runtime. This indicates that a next step in future work towards improving the proposed method might possibly be to try to optimize the choosing of the colliding pair of molecules, e.g.\ by improving on the data structures storing the state of the system.

\subsection{Cache as a partial record of network}

At the end of a simulation, the cache contains the most recently used reaction instances, effectively storing a subnetwork of the explored reaction space.
More precisely, as described in Section~\ref{cacheScheme}, each element of the cache stores a list of all reaction instances possible to generate by applying one template $r$ to one molecular pair $(m_1, m_2)$. It is the union of those lists that we consider the subnetwork represented by the cache.
Thus, while the stochastic simulation itself does not require explicit construction of the full reaction network, the cache provides a controllable, interpretable and lightweight representation of a portion of that network.

\begin{figure}[t]
    \centering
    \includegraphics[width=0.75\textwidth]{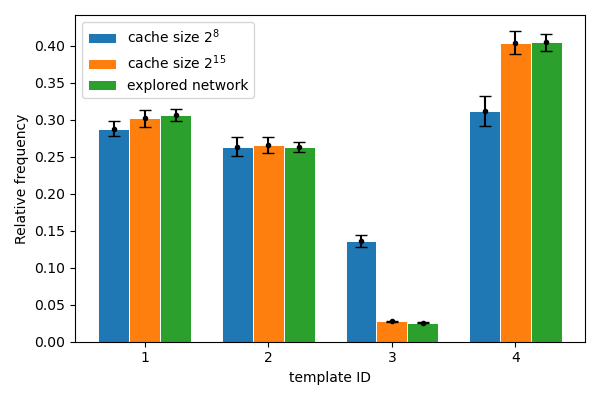}
    \caption{Comparison of the relative distribution of the four reaction templates among the successful applications of a reaction template~$r$ on a molecular pair $(m_1,m_2)$. Successful means that the left-hand side of $r$ has one or more matches as a subgraph of $m_1 \cup m_2$. The blue bars show this for the contents of a cache of size~$2^8$ at the end of the simulation, the orange bars show the same for a cache of size~$2^{15}$, and the green bars show the distribution across all steps of the entire simulation. The error bars represent the standard deviation in the frequency over the $10$ trials conducted.}
    \label{fig:frequencyTemplate}
\end{figure}

By modifying the cache size or replacement policy, different regions of the explored network can be preferentially retained without altering the stochastic trajectory of the simulation. In the extreme case of a sufficiently large cache, all reaction instances encountered during the simulation would be retained, allowing reconstruction of the complete reaction network explored by the stochastic process. Therefore, the cache can be considered as a \emph{proxy} of the molecular network explored by the simulation. In this manner, the cache has a practical utility and a conceptual contribution to the analysis of the results of the simulation besides speeding it up. It might allow the possibility of analysing properties of the network without storing the entire enumerated molecular network.

To investigate to what extent the cached portion is representative of the entire explored network, we for caches of two sizes ($2^8$ and $2^{15}$) compared the coverage of reactions and molecules in their final state with that of the entire simulation.

Figure~\ref{fig:frequencyTemplate} shows the relative distribution of the four reaction templates among the successful applications of a reaction template~$r$ on a molecular pair $(m_1,m_2)$. Successful means that the left-hand side of $r$ has one or more matches as a subgraph of $m_1 \cup m_2$. The blue bars show this distribution for the keys stored in a cache of size~$2^8$ at the end of a simulation of $10^6$ steps (the keys of the cache have type $(r, m_1,m_2)$ and successful is equivalent to the list of the cache element being non-empty). The orange bars show the same distribution for a cache of size~$2^{15}$. The green bars show the same distribution for all steps of the entire simulation (but only counting the first successful application for each $(r, m_1,m_2)$ triple, not repetitions, to be consistent with the counting of the cache contents).
The error bars depict the standard deviation over $10$ trials of the simulation.
The relative ordering of the frequencies of the reaction templates (which result in reaction instances) remains the same in all three cases, with cache of size $2^8$, $2^{15}$ and the explored network, namely template $4$ is the most frequent, followed by $1$, $2$ and $3$. The absolute counts of the number of times the particular template appears in the cache with instantiated reaction differ slightly.
Reactions generated using template $2$ are overrepresented in the cache of size $2^8$ at the expense of reactions effected using template $3$. 
Using a cache of size $2^{15}$, the frequencies of the templates closely follow the frequencies of the templates instantiated in the entire explored network.

\begin{table}
	\centering
	\begin{tabular}{@{}S[table-format = > 1]S[table-format=2]*{2}{S[table-format=3]}@{}}
	\toprule
	{Number of carbon atoms} & {Cache size $2^8$} & {Cache size $2^{15}$} & {Explored network} \\
	\midrule
	1 & 1 & 1 & 1 \\
	2 & 2 & 2 & 2 \\
	3 & 3 & 3 & 3 \\
	4 & 5 & 5 & 5 \\
	5 & 9 & 9 & 9 \\
	6 & 17 & 17 & 17 \\
	7 & 30 & 34 & 34 \\
	> 7 & 48 & 426 & 431 \\
	\bottomrule
    \end{tabular}
    \caption{A comparison of the number of classes of molecules with different number of carbon atoms in the two caches with different sizes.}
    \label{fig:molSizes}
\end{table}

Table~\ref{fig:molSizes} lists the number of molecular classes with increasing number of carbon atoms in the cached subnetwork, when cache size is $2^8$ and $2^{15}$, and the entire explored network, respectively. 
All the molecular classes with up to $6$ carbon atoms from the explored molecular network were present in both the caches. The molecular classes which are rarer and have more carbon atoms are the ones not present in the cache with size $2^8$: the cache contains only $48$ molecules with more than $7$ carbon atoms while there were $431$ such molecules in the entire explored network. The cache of size $2^{15}$ could capture $426$ of these $431$ molecules. Among the $115$ molecular classes in the cache with size $2^8$, $97\pm1$ of the $100$ most abundant molecular classes in the entire network were present. Out of the $100$ most frequently queried molecular pair-reaction template combinations during the simulation, $94\pm2$ were present in the cached subnetwork (with minor variations over trial runs). Thus, it was the rare molecular pair-reaction template queries that were absent in the cache, which is what would be expected from an efficient cache aiming to maximize the hit rate during the simulation.  

\section{Summary and Outlook}

We have presented a framework for exploring molecular reaction networks, motivated by chemical stochastic kinetics generating random walks on networks~\cite{randomwalk1,randomwalk2,randomwalk3} that differ from previous approaches. We neither do not precompute the network but rather discover regions of it through a process mimicking chemical stochastic dynamics. Instead of computing the relative probabilities for all possible reactions at each step, our method performs three modular sampling operations: (1) selecting a colliding molecular pair based on current molecular abundances, (2) selecting a reaction template, and (3) selecting a specific reaction instance generated by a graph transformation engine acting on that pair. In the study presented here, the reaction template probabilities are not intended to approximate physical rate laws, but rather to provide an unbiased sampling of dynamically reachable regions of a reaction graph defined on an atomistic level. This divide-and-conquer strategy avoids enumerating the full list of transitions out of the current state at every step and allows the dynamics to generate previously unvisited regions of chemical space on demand. In this sense, the method performs a biased random walk on an implicitly defined molecular network, constructing only the locally relevant subnetworks during the course of the exploration. 

Some key features of the framework proposed in the current work include the simultaneous use of (1) stochastic simulation-styled dynamics operating on molecular graphs (2) applying rule-based bond rearrangements on an atomic level using reaction templates to generate a relevant part of the molecular network on the fly and (3) the support of open systems with inflow and outflow on the system. 
It borrows from and straddles the region between the Gillespie algorithm for stochastic simulations (which is kinetically accurate but abstract), 
generative chemistry using graph grammar (which is chemically accurate but deterministic) and kinetic Monte Carlo simulations (which are stochastic but non-generative and closed). 
The simulation can be steered to explore the region of the molecular space under interest--- its parameters might also be biased to explore events which rarely occur under realistic conditions but are nevertheless important.
The present method provides a lightweight, highly scalable way to navigate regions of chemical space where kinetic parameters are unknown or where the objective is discovery rather than quantitative prediction.
It enables stochastic, network-free exploration of large, generative molecular spaces under minimal kinetic assumptions.
Together, with stochastic simulation schemes like the Gillespie algorithm or kinetic Monte Carlo, it forms two complementary probes to molecular networks: one for exact stochastic dynamics, one for graph-based exploration.

Several directions exist for improving the physical fidelity and computational efficiency of the framework:
\begin{enumerate}
    \item \emph{Kinetically informed sampling}: The two random-choice steps for selecting reaction templates and instances are intentionally simplified. A more realistic dynamics would incorporate reaction probabilities derived from the law of mass action, which requires rate constants (requiring estimates of activation energy barriers). Although computing these quantities is challenging, incorporating them would be necessary for recovering quantitatively accurate kinetics rather than purely exploratory behaviour. We expect caching reaction instances along with their energy barriers would significantly reduce the computation and speed up the simulation. Most simulation algorithms utilize educated guesses for the rate constants, which makes the behaviour of the system particularly sensitive to the inputs for these parameters. 
    \item \emph{Free Energy difference weighted instance selection}:
    	As a less ambitious alternative, reaction instance selection could use a Metropolis Hastings type acceptance rule (which is often used in high dimensional systems).
    	We sketch a possible scheme using weights of the form $\exp(-E_i/(\alpha k_B T))$ in Figure~\ref{fig:reac_mh_prob}, where $E_i$ is the free energy difference associated with the products of the $i$th candidate reaction and $\alpha$ is a parameter to adjust the relative probabilities of the candidate reactions. While such weights resemble those used in equilibrium sampling (e.g., grand canonical ensembles) in thermodynamic Monte Carlo simulations, adapting them to chemical kinetic transitions may offer a controllable bias with reaction pathways towards energetically favourable regions of the network. Moreover, in simulations to elucidate the equilibrium state through Monte Carlo simulations, the energy $E_i$ used in the acceptance rule is the free energy of the sampled state, and not the difference in free energies of two states.

    \begin{figure}
        \centering
        \includegraphics[width=0.5\linewidth]{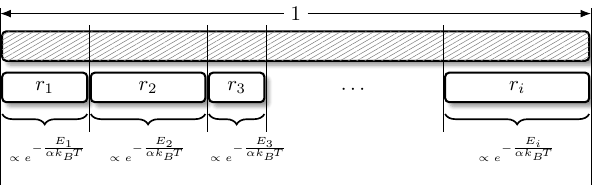}
        \caption{The calculation of the weights using a Metropolis-Hastings-type scheme after applying a reaction template to the selected pair of molecules generating the reactions $\{r_1, r_2, r_3, \cdots r_i\}$.}
        \label{fig:reac_mh_prob}
    \end{figure}
    
    \item \emph{Pathway tracing and caching strategies}: Many scientific questions, such as identifying plausible formation pathways from input molecules for queried molecules, require keeping track of reaction provenance. Alternative caching strategies can be investigated to quantify the speed-up they provide to the simulation and select the subspace of the implicit network explored by the algorithm. More sophisticated caching, indexing, or provenance-tracking structures could be developed depending on the needs of the study.
    \item \emph{Spatio-temporal simulations}: The framework can be naturally extended to simulate systems that are not well mixed by partitioning the volume or surface into multiple coupled regions (possibly arranged on a lattice). Each region hosts an independent instance of the stochastic simulation, while diffusion is modelled as inflow-outflow events between adjacent regions. This enables the study of systems with spatial heterogeneity emerging from the coupling of stochastic kinetics with diffusive transport. This allows investigation of concentration gradients, travelling waves, and oscillations, which arise in chemical systems with feedback loops and nonlinear kinetics.
    
    Coupling generative, rule-based frameworks with spatial discretization allows local differences in the exploration of molecular space while permitting mass exchange between them. This opens the possibility of studying how spatial constraints influence molecular diversity, and the emergence of phenomena that are difficult to capture in well-mixed models: localized accumulation and shielding of molecules from dilution, or enabling compartmentalized autocatalysis. This broadens the applicability of the framework to heterogeneous environments on films, membranes, microfluidics, and early-earth or extraterrestrial settings.
\end{enumerate}
Future work will focus on incorporating these refinements in order to increase both accuracy and scalability. By combining template-based modelling with stochastic exploration, the framework opens a path toward systematically probing large chemical spaces that are otherwise difficult to enumerate or simulate with conventional deterministic or kinetics-driven methods. 

\section*{Acknowledgement}

This work was supported by funding from the European Union's Horizon 2021 Research
and Innovation program under Marie
Sklodowska-Curie
grant agreement no.~101072930  (TACsy --- Training Alliance for Computational systems chemistry).

\bibliographystyle{unsrt}
\bibliography{sample}

\appendix

\section{Cache Replacement Policy}
\label{crp}

The cache streams in reaction instances until it is full, and thereafter overwrites existing reaction instances with incoming ones according to a cache replacement policy. The optimal cache replacement algorithm OPT~\cite{cache_opt} would overwrite the reaction instances that will not be requested for the largest number of simulation steps in the future. However, it is difficult to predict after how many steps a reaction would be requested again, in particular in a stochastic simulation.

\begin{figure}[!bh]
    \centering
    \begin{subfigure}{0.30\textwidth}
        \centering
        \includegraphics[scale=0.75]{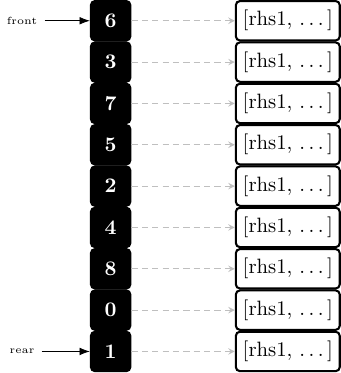}
        \caption{Depiction of the ordered dictionary used as the cache data structure.}
        \label{fig:cache}
    \end{subfigure}%
    \hfill
    \begin{subfigure}{0.32\textwidth}
        \centering
        \includegraphics[scale=0.75]{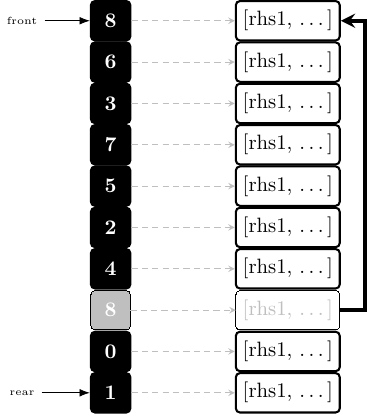}
        \caption{Cache state after a cache hit moves the requested element to the front.}
        \label{fig:cache_hit}
    \end{subfigure}%
    \hfill
    \begin{subfigure}{0.32\textwidth}
        \centering
        \includegraphics[scale=0.75]{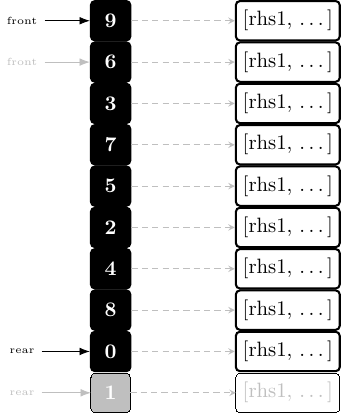}
        \caption{Cache state after a cache miss removes the rear element and inserts the new element at the front.}
        \label{fig:cache_miss}
    \end{subfigure}
    \caption{The LRU caching policy implemented via an ordered dictionary. A key in the dictionary is formed by a canonical representation of two molecular graphs $(m_1,m_2)$ and the index of the reaction template $r$, while each value is a list containing the right hand sides of the reaction instances computed by applying all matches found between the left hand side of $r$ and the union of the molecular graphs $m_1\cup m_2$. For brevity, the keys are represented by single digit numbers, and the values by abstracted lists. Note that a value might be null for some keys, if the left hand side of $r$ does not match $m_1\cup m_2$. A cache hit results in reordering the entries in the dictionary, by moving the matched query to the front of the dictionary as shown in Figure~\ref{fig:cache_hit}. On a cache miss, the least recently used element (which is the element at the rear of the dictionary) is removed and the new element is added to the front of the dictionary as shown in Figure~\ref{fig:cache_miss}.}
    \label{fig:cache_model}
\end{figure}

Two simple cache replacement policies without significant overhead costs of tracking the access history of cache elements are random replacement and first-in first-out queue. These require minimal bookkeeping but often result in poor hit rates. The cache replacement policy used in this work is that the \emph{least recently used} (LRU) entry is overwritten. This policy assumes that reaction instances that were more recently used in the past would be requested closer in the future. We maintain what in Python is called an \emph{ordered dictionary}~\cite{ordered_dict_py} (essentially a linked list whose elements can be accessed via a hash function) containing the already computed reaction instance groups. In the (key,value)-pairs of the dictionary, a key is composed of a molecule pair $(m_1,m_2)$ and a reaction template $r$, and the value is a list of all the reactions possible to generate by applying $r$ to $(m_1,m_2)$. The most recently used cache element is the first entry of the ordered dictionary, while the least recently used element is the last, as shown in Figure~\ref{fig:cache}. Note that the choice of keys fits the working of Algorithm~\ref{generate_network}, which during the generation of a candidate transition for a state exactly has the colliding pair of molecules and the selected reaction template available (whereas it has no knowledge of their time of last use). On a cache hit, the requested cache element is retrieved from the ordered dictionary and moved to the front of the ordered dictionary, since it now is the most recently used cache element. This is shown in Figure~\ref{fig:cache_hit}. On a cache miss, the least recently used element, which is at the end of the dictionary, is removed. The new element is pushed to the front of the ordered dictionary, as shown in Figure~\ref{fig:cache_miss}. This ensures that the elements in the dictionary are ordered by the the number of steps lapsed since they were requested last. Operations of an ordered dictionary have expected $\mathcal{O}(1)$ runtime (as follows from the runtimes of the underlying hash table and linked list), hence the cache operations have that same runtime.

This algorithm can be shown to result in at most $N$ times more cache misses than OPT (where $N$ is the size of the cache)~\cite{lru_cache_performance}. The worst case occurs when $N+1$ queries are accessed sequentially in an iterative manner. The $(N+1)$th data element overwrites the first element which was the least recently used element. This pattern continues in all subsequent iterations, which therefore all will result in cache misses. These cyclic access patterns, though rare, might occur during the stochastic simulation. In those cases the cache replacement policy might be improved by overwriting the cache element whose $k$-th last request was furthest in the past. Another alternative is to keep track of the frequency of requests of each cache element and overwrite the least \emph{frequently} requested element on a cache miss. However, keeping a count of the request frequency of each cache element would require a third data structure for the necessary bookkeeping, which is not as straight-forward as ordering the cache elements in the linked list by their recency of requests. 
Also, older cache elements would tend to have more requests than recent ones, leading to the more recent cache elements being overwritten. A frequency based caching policy would retain such non-recent cache elements with higher frequency of access in contract to a recency of use based policy.

\section{List of Molecular Structures}

\begin{figure}[h]
    \centering
    \includegraphics[width=0.5\linewidth]{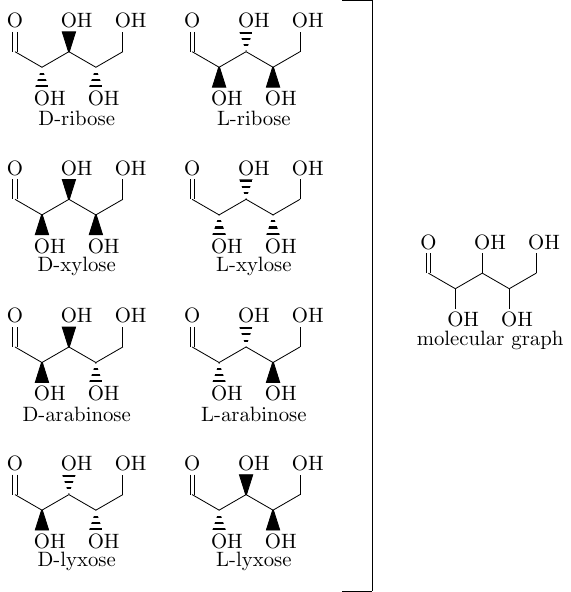}
    \caption{Stereoisomers are represented by the same molecular graphs, leading to them being indistinguishable in the simulation.}
    \label{fig:stereoisomers}
\end{figure}

As mentioned in Section~\ref{exp}, the molecular graphs used in the simulation do not encode stereochemistry. Therefore, stereoisomers (which are important while considering carbohydrate molecules) are represented by a single molecular graph as shown in Figure~\ref{fig:stereoisomers}.

Table~\ref{tab:5CMols} lists all the molecular graphs with five carbon atoms that were generated by the simulation, along with their SMILES strings and the index used as labels while referring to molecules with five carbon atoms
(in Figures~\ref{fig:5cDistNoFlow},~\ref{fig:5cDistFlow} and~\ref{fig:rn5c}).
Two out of the eight stereoisomers represented by the molecular graph indexed $1$ correspond to the sugar ribose.

Similarly, Table~\ref{tab:6CMols} lists all the molecular graphs with six carbon atoms that were generated by the simulation, along with their SMILES strings and the index used as labels while referring to molecules with six carbon atoms
(in Figures~\ref{fig:6cDistNoFlow},~\ref{fig:6cDistFlow} and~\ref{fig:rn6c}).
Two out of the sixteen stereoisomers represented by the molecular graph indexed $1$ correspond to the sugar glucose, while two out of the sixteen stereoisomers represented by the molecular graph indexed $6$ correspond to the sugar fructose.

\begin{table}[h]
    \centering
    \begin{tiny}
        \begin{tabular}{rcc}
            index & smiles & structure \\\hline
            \rowcolor{mygray} & & \\
            \rowcolor{mygray} 1 & C(C(C(C(CO)O)O)O)=O & \chemfig{HO-[:30]-[:-30](-[:-90]OH)-[:30](-[:90]OH)-[:-30](-[:-90]OH)-[:30]=[:90]O} \\\\
            2 & C(O)=C(C(C(CO)O)O)O & \chemfig{HO-[:30]-[:-30](-[:-90]OH)-[:30](-[:90]OH)-[:-30](-[:-90]OH)=[:30]-[:90]OH} \\\\
            3 & C(C(CO)(C(CO)O)O)=O & \chemfig{HO-[:30]-[:-30](-[:-90]OH)-[:30](-[:120]HO)(-[:60]-[:-30]OH)-[:-30](=[:-90]O)} \\\\
            4 & C(C(CO)O)(C(CO)O)=O & \chemfig{HO-[:30]-[:-30](-[:-90]OH)-[:30](=[:90]O)-[:-30](-[:-90]OH)-[:30]-[:90]OH} \\\\
            5 & C(CO)(C(C(CO)O)O)=O & \chemfig{HO-[:30]-[:-30](-[:-90]OH)-[:30](-[:90]OH)-[:-30](=[:-90]O)-[:30]-[:90]OH} \\\\
            6 & C(C(CO)O)(O)=C(CO)O & \chemfig{HO-[:30]-[:-30](-[:-90]OH)-[:30](-[:90]OH)=[:-30](-[:-90]OH)-[:30]-[:90]OH} \\\\
            7 & C(C(C(CO)(CO)O)O)=O & \chemfig{O=[:30]-[:-30](-[:-90]OH)-[:30](-[:120]HO)(-[:60]-[:-30]OH)-[:-30](-[:-90]OH)} \\\\
            8 & C(O)=C(C(CO)(CO)O)O & \chemfig{HO-[:30]=[:-30](-[:-90]OH)-[:30](-[:120]HO)(-[:60]-[:-30]OH)-[:-30](-[:-90]OH)} \\\\
        \end{tabular}
    \end{tiny}
    \caption{Structures of the five molecules with five carbon atoms in the system with their SMILES strings which are used as labels for the ticks on the y-axis in Figure~\ref{fig:5cDistNoFlow} and Figure~\ref{fig:5cDistFlow}}
    \label{tab:5CMols}
\end{table}

\begin{table}[h]
    \centering
    \begin{tiny}
        \begin{tabular}{rcc}
            index & smiles & structure \\\hline
            \rowcolor{mygray} & & \\
            \rowcolor{mygray} 1 & C(C(C(C(C(CO)O)O)O)O)=O & \chemfig{HO-[:30]-[:-30](-[:-90]OH)-[:30](-[:90]OH)-[:-30](-[:-90]OH)-[:30](-[:90]OH)-[:-30]=[:-90]O} \\\\
            2 & C(O)=C(C(C(C(CO)O)O)O)O & \chemfig{HO-[:30]-[:-30](-[:-90]OH)-[:30](-[:90]OH)-[:-30](-[:-90]OH)-[:30](-[:90]OH)=[:-30]-[:-90]OH} \\\\
            3 & C(C(C(CO)O)(C(CO)O)O)=O & \chemfig{HO-[:30]-[:-30](-[:-90]OH)-[:30](-[:120]HO)(-[:60]=[:-30]O)-[:-30](-[:-90]OH)-[:30]-[:-30]OH} \\\\
            4 & C(C(C(CO)O)O)(C(CO)O)=O & \chemfig{HO-[:30]-[:-30](-[:-90]OH)-[:30](-[:90]OH)-[:-30](=[:-90]O)-[:30](-[:90]OH)-[:-30]-[:-90]OH} \\\\
            5 & C(C(CO)(C(C(CO)O)O)O)=O & \chemfig{HO-[:30]-[:-30](-[:-90]OH)-[:30](-[:90]HO)-[:-30](-[:-90]OH)(-[:-30]=[:30]O)-[:30]-[:90]OH} \\\\
            \rowcolor{mygray} 6 & C(CO)(C(C(C(CO)O)O)O)=O & \chemfig{HO-[:30]-[:-30](-[:-90]OH)-[:30](-[:90]OH)-[:-30](-[:-90]OH)-[:30](=[:90]O)-[:-30]-[:-90]OH} \\\\
            7 & C(C(C(CO)O)O)(O)=C(CO)O & \chemfig{HO-[:30]-[:-30](-[:-90]OH)-[:30](-[:90]OH)-[:-30](-[:-90]OH)=[:30](-[:90]OH)-[:-30]-[:-90]OH} \\\\
            8 & C(C(CO)O)(O)=C(C(CO)O)O & \chemfig{HO-[:30]-[:-30](-[:-90]OH)-[:30](-[:90]OH)=[:-30](-[:-90]OH)-[:30](-[:90]OH)-[:-30]-[:-90]OH} \\\\
            9 & C(C(C(CO)(C(CO)O)O)O)=O & \chemfig{HO-[:30]-[:-30](-[:-90]OH)-[:30](-[:120]HO)(-[:60]-[:-30]OH)-[:-30](-[:-90]OH)-[:30]=[:-30]O}
        \end{tabular}
        \hspace{1cm}
        \begin{tabular}{rcc}
            index & smiles & structure \\\hline\\
            9 & C(C(C(CO)(C(CO)O)O)O)=O & \chemfig{HO-[:30]-[:-30](-[:-90]OH)-[:30](-[:120]HO)(-[:60]-[:-30]OH)-[:-30](-[:-90]OH)-[:30]=[:-30]O} \\\\
            10 & C(C(C(C(C(CO)O)O)O)O)=O & \chemfig{HO-[:30]-[:-30](-[:-60]OH)(-[:-120]-[:180]OH)-[:30](-[:120]OH)(-[:60]-[:-30]HO)-[:-30]=[:-90]O} \\\\
            11 & C(O)=C(C(C(C(CO)O)O)O)O & \chemfig{HO-[:30]-[:-30](=[:-90]O)-[:30](-[:90]HO)-[:-30](-[:-90]OH)(-[:-30]-[:30]OH)-[:30]-[:90]OH} \\\\
            12 & C(C(C(CO)O)(C(CO)O)O)=O & \chemfig{O=[:30]-[:-30](-[:-90]OH)-[:30](-[:90]HO)-[:-30](-[:-90]OH)(-[:-30]-[:30]OH)-[:30]-[:90]OH} \\\\
            13 & C(C(C(CO)O)O)(C(CO)O)=O & \chemfig{HO-[:30]=[:-30](-[:-90]OH)-[:30](-[:120]HO)(-[:60]-[:-30]OH)-[:-30](-[:-90]OH)-[:30]-[:-30]OH} \\\\
            14 & C(C(CO)(C(C(CO)O)O)O)=O & \chemfig{HO-[:30]=[:-30](-[:-90]OH)-[:30](-[:90]HO)-[:-30](-[:-90]OH)(-[:-30]-[:30]OH)-[:30]-[:90]OH} \\\\[0.1cm]
            15 & C(CO)(C(C(C(CO)O)O)O)=O & \chemfig{HO-[:30]-[:-30](-[:-90]OH)=[:30](-[:90]HO)-[:-30](-[:-90]OH)(-[:-30]-[:30]OH)-[:30]-[:90]OH} \\\\
            16 & C(C(C(CO)O)O)(O)=C(CO)O & \chemfig{HO-[:30]-[:-30](=[:-90]O)-[:30](-[:120]HO)(-[:60]-[:-30]OH)-[:-30](-[:-90]OH)-[:30]-[:-30]OH} \\\\
            17 & C(C(CO)O)(O)=C(C(CO)O)O & \chemfig{HO-[:30]-[:-30](-[:-90]OH)-[:30](=[:90]O)-[:-30](-[:-90]OH)(-[:-30]-[:30]OH)-[:30]-[:90]OH} 
        \end{tabular}
    \end{tiny}
    \caption{Structures of the seventeen molecules with six carbon atoms in the system with their SMILES strings which are used as labels for the ticks on the y-axis in Figure~\ref{fig:6cDistNoFlow} and Figure~\ref{fig:6cDistFlow}.}
    \label{tab:6CMols}
\end{table}

\end{document}